# Estimation of Elastic Behavior of Cast Metal Components Containing Process Induced Porosity

Shiguang Deng [a, *], Carl Soderhjelm [a], Diran Apelian [a], Krishnan Suresh [b]

[a] ACRC, Materials Science and Engineering, University of California, Irvine, CA, USA
[b] Mechanical Engineering, University of Wisconsin, Madison, WI, USA

**Abstract**

Significant progress has been made for assessing the influence of porosity on the performance metrics for cast components through various modeling techniques. However, a computationally *efficient* framework to account for porosity with various shapes and sizes is still lacking. The main contribution of this work is to address this limitation. Specifically, a novel porosity sensitivity method is proposed, which integrates the merits of topological sensitivity and shape sensitivity. While topological sensitivity approximates the first order change on the quantity of interest when an infinitesimally small spherical pore is inserted into a dense (no pore) structure, shape sensitivity estimates the subsequent change in the quantity when the small pore's boundary is continuously perturbed to resemble the geometry reconstructed from tomography characterization data. In this method, an exterior problem is solved to explicitly formulate pore stress and strain fields as functions of shape scaling parameters. By neglecting higher order pore-to-pore interaction terms, the influence of multiple pores can be estimated through a linear approximation. The proposed method is first studied on a benchmark example to establish the impact of different pore parameters on the estimation accuracy. The method is then applied onto case studies where the pore geometry is either from tomography reconstruction or computer-generated representations. Efficiency and accuracy of the method are finally demonstrated using a commercial 3D application. The proposed method can be extended to other manufacturing (e.g., additive manufacturing) induced porosity problems.

## 1. INTRODUCTION

Cast aluminum alloys are widely used for structural automobile components manufactured by high pressure die casting (HPDC) [1]–[5]. HPDC is a quick, reliable, economic process with the capability to manufacture high-volume net-shaped components with tight tolerance. However, a major concern for cast components is the presence of defects, the main one being porosity. More specifically, in the cold chamber HPDC process [6]–[8], entrained air along with molten metal is injected into the mold at high speed, which if not removed prior to solidification will result in gas porosity. In addition to entrained gas porosity, shrinkage porosity occurs during solidification in regions which solidify slower and may suffer from inadequate liquid metal flow from the surrounding metal that has solidified [9]. Presence of either type of porosity will impair the mechanical performance of cast components; however, it is difficult to represent performance degradation in a quantitative manner as a function of pore characteristics.

The influence of porosity can be studied via several different approaches as illustrated in Figure 1. Early work involved the development of analytical upper and lower bounds for elastic moduli of multiphase materials [10]–[12]. When the ratio between different phase moduli is relatively small, derived bounds estimate well the homogenized effective moduli. However, for multiphase materials with high contrast phase properties, the gap between bounds can be fairly large. In extreme scenarios, the lower bound could deteriorate to zero [10], [13], making it unsuitable for analysis; much of the analysis to date has been based on heuristic models [14]–[16]. The influence of pore size, morphology, spatial ramification, and topological distribution on the mechanical behavior of HPDC cast components is the focus of this work.

Computed tomography (CT) allows us to characterize the spatial characteristics of the porosity present in cast components. Geometrical features of pores are then reconstructed in 3D models through image processing of the scanned data [13], [17]–[19]. Porosity detection relies on the resolution of tomography scanner, which could vary from sub-micron to sub-millimeter. Knowledge of the minimum pore size will influence the choice of the resolution of the tomography scanner needed [13], [19].

With CT as an enabling technology, one alternative to the analytical bound approach [10] is through numerical simulations. Previous work has incorporated tomography reconstructed geometries into a finite element (FE) model, followed by a direct FE analysis. Actual shape and size of casting pores were characterized by light microscopy and reproduced in a FE model to correlate with local stress concentrators [20]. A micromechanical model incorporating pore geometries reconstructed from tomography has been developed to predict the influence of porosity on elastoplastic behavior [21]. Shan and Gokhale [22] incorporated quantitative description of non-uniformly distributed pores into 3D micromechanical analysis of materials. It was found that the distribution of local stresses and strains depends on size, orientation, and spatial arrangement of pores. Gas and shrinkage induced pores were distinguished by fractal analyses in terms of individual shapes and congregated spatial distributions [23]. While such methods were directly built on pore geometries, several numerical issues can arise. Incorporating pore characteristics increases computational time and complexity of FE meshing and solution processes. Specifically, smaller elements are generated in the vicinity of pore surfaces for geometry adaptivity and element size transition. This results in substantial increase of mesh size and element quality issues, such as distorted or inverted elements [24]. Ill-shaped elements deteriorate the condition number of global stiffness matrix in the FE system, leading to slow convergence or even failure [25].

---

*Corresponding author.
Email address: sdeng9@wisc.edu (Shiguang Deng)

Even though advances in multiscale FE methods and synthetic porosity models have been made over the last decade, quantitatively correlating porosity and part performance is complex. Typically, deformation gradient is provided from macroscale to local microstructures with computer generated simplified porosity models. The deformation gradient is then transformed to micromechanics boundary value problems, which allows us to solve local homogenized stresses, and this in turn is fed back into macrolevel models. Temizer and Wriggers [26] developed a tangent information-based condensation procedure to minimize the number of micromechanical testing to enforce deformation-controlled boundary conditions via perturbation methods. Taxer et al. [13] proposed a sequential multiscale model to account for the influence of casting porosity on the plastic behavior of Ni based superalloys. Macroscopic regions were divided into no-pore and porous regions in which porosity dependent local material constitutive properties were applied. The challenge in such multilevel models is the high memory allocation costs [27]. Another challenge comes from the fact that in order to statistically account for porosity property distributions, it is necessary to extract a large number of physical samples from castings and examine each sample using tomography [13], which is labor intensive, time consuming and cost-ineffective. As stated earlier, it is a challenging and complex task to quantitatively correlate porosity and part performance.

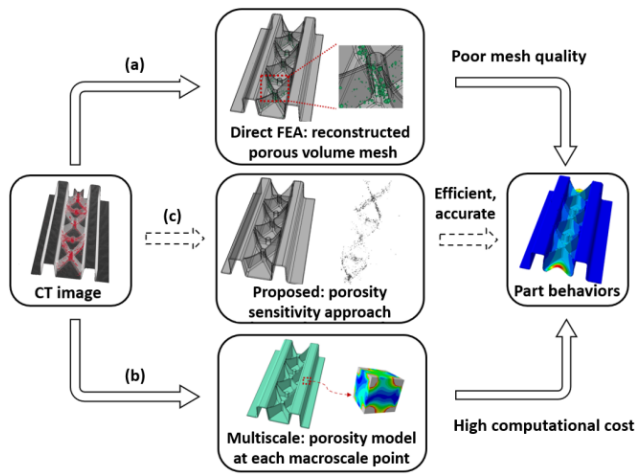

*Figure 1*: Overview of different approaches to analyze porous metal structures: **(a)** the direct FEA approach requires tomography reconstructed porous volume mesh whose high computational expense is due to a large number of severely distorted elements; **(b)** the classic multiscale simulation approach needs to characterize pore distributions on each macroscale integration point (IP) and its high computational cost comes from the frequent data exchange between scales; **(c)** the proposed computationally efficient porosity sensitivity approach.

Defeaturing is a model simplification technique originating from the field of computer-aided-design (CAD) to preserve certain model properties due to geometrical variations [28]. Shape sensitivity as well as topological sensitivity have shown to be useful in such computations. Shape sensitivity computes the change of performance when the model boundary is infinitesimally perturbed [29]; whereas topological sensitivity captures the change when an infinitesimally small hole is created in the domain [30]. Li et al. [31] estimated and accounted for the error of suppressing any arbitrarily sized negative or positive geometry features using adjoint theory. In their later work [32], a second-order defeaturing method was proposed to account for structural performance change when multiple interacted geometry features were suppressed. Gopalakrishnan and Suresh [33] quantified the bounds of defeaturing induced error through monotonicity analysis. However, under certain circumstance, the bounded interval became too large to be useful. Turevsky et al. [34] developed a feature sensitivity approach by integrating shape sensitivities over shape transformations to formulate defeaturing errors for 2D Poisson problems where topological change was estimated by shape sensitivity. Deng and Suresh [35] proposed a topological sensitivity based metric to predict the potential benefits for topology optimization.

Despite these advances, gaps exist to quantitatively describe FEA models that accurately and efficiently portray the complexities of defects such as porosity in cast components; specifically:

- Direct FE approaches utilizing tomography reconstructed pore geometry have mesh quality and convergence issues, while multiscale models are computationally cost-ineffective. There is a need for a computationally efficient framework to correlate pore shape, size and spatial characteristics with casting performance.
- Defeaturing techniques have mostly been applied to 2D simple problems with few geometrical features [34]. They fall short in describing linear elastic quantities for randomly distributed pores in 3D.
- Topology and shape sensitivities are mainly utilized in structural optimizations to provide derivative information for mathematical programming [36], [37]. To our best knowledge, this work is the first to use the two design sensitivities to account for the influence of manufacturing porosity on part performances.

We propose a first-order porosity sensitivity-based estimator to predict the impact of arbitrarily shaped porosity on casting behavior. In this method, the porous structure is divided into two interconnected subdomains: a dense structure without pores, and a porosity domain. The efficacy of the proposed method is two-fold. First, without irregularly shaped pores, the mesh size and quality of the dense structure can be much improved, leading to fast convergent linear systems. Second, the porosity sensitivity derived in this work is formulated in boundary representation which only requires 2D mesh on pore surfaces. Compared with 3D volume elements in a traditional FE analysis, computing on 2D surfaces is more efficient.

The remainder of the paper is organized as follows. Section 2 describes material characterizations and pore properties. Section 3 mathematically and quantitatively describes the problem. Section 4 reviews technical background on topological and shape sensitivities. Section 5 describes the proposed method. In Section 6, the robustness and accuracy of the proposed method are evaluated through numerical experiments. Finally, Section 7 provides additional discussion and conclusions.

## 2. POROSITY CHARACTERIZATION

The material of interest in this study is an aluminum based alloy which was produced at Magna International [38] via vacuum assisted HPDC [4]. The sample was produced in the shape of W-profile plate and the casting module is shown in Figure 2(a) where the cross-sectional thickness is 3 mm. Additional details of part dimensions are given in Figure 39. It should be noted that for this study, the sample casting was purposely cast with excessive porosity. The sample underwent a stress relief and solutionizing heat treatment for 75 minutes at 460 °C followed by a forced air quench at the rate of 40 °C per second, followed by an artificial age heat treatment for 108 minutes at 215 °C.

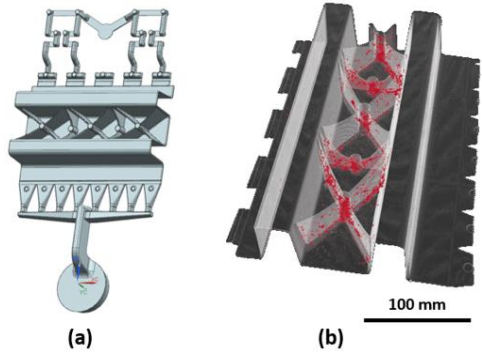

*Figure 2: (a) Casting module for W-profile plate with feeder, runner, and gate; (b) tomography reconstructed volume with pores distributed within the casting.*

Porosity in the cast component was characterized via computed tomography (CT) at VJT's high energy X-ray laboratory [39] using a linear accelerator (LINAC) X-ray source with 6 MV penetration energy. The sample was scanned in a 'step and shoot' mode; 3 frames of images were shot per step. While the first frame was to remove any possible device vibrations and 'burn in' effects, the latter two steps were averaged for reconstruction. In total, 2400 steps were taken over $360^0$. While each frame took 5000 microseconds, the entire scanning was carried out in 105 minutes. CT characterization has a resolution of 180 μm per voxel. The acquired radiograph was converted to a binary image and transferred to a scanning data processing software, VOLEX [40], for tomographic reconstruction, where a reliable pore morphology description required a 2X2X2 surface connected voxel filter. Selection of this filter value resulted in the smallest detectable pore with the size of 500 μm. For this W-profile plate, its volume (362,758 mm$^3$) is represented by 61,978,130 voxels and its longest dimension (295 mm) is discretized by 1,639 voxels. The volume of detected pore is 1,109 mm$^3$, accounting for 0.3% of total material volume. The tomography reconstructed volume containing detected pores is shown in Figure 2(b).

The degree of porosity detected is largely determined by CT resolution. While a low resolution could generate light-weight image graphs by only capturing large size pores, it cannot detect small pores which are prevalent in cast parts. In order to account for various degrees of porosity and for the sake of computational cost, we considered pore whose equivalent diameter is larger than 500 μm. To further quantify pore characteristics, each pore is represented by four descriptors: equivalent diameter, equivalent sphericity, distance to the nearest pore neighbor, and distance to part surfaces. Their cumulative distributions are shown in Figure 3. Sphericity measures how close the shape of a pore is to that of a perfect sphere; perfect sphere having a value of 1.0.

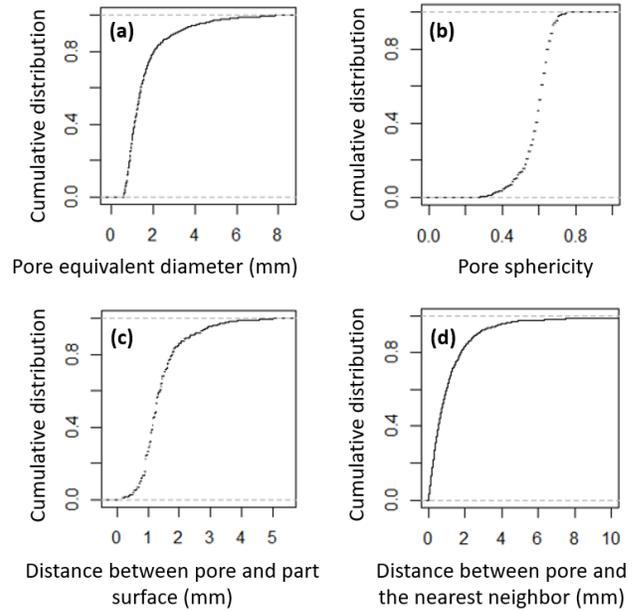

*Figure 3: Porosity data from tomography reconstruction on W-profile plate: (a) cumulative distribution of pore equivalent diameter, (b) pore sphericity, (c) pore-to-surface distance, and (d) pore-to-pore distance.*

As can be seen from Figure 3, the median value of pore sphericity is 0.7. The median value of pore equivalent diameter is 1.4 mm. The median distance between pore and part surface is 1.5 mm. The median value of distances between pores and their nearest neighbors is 1.2 mm. While gas induced pores tend to have a high sphericity, solidification induced pores tend to have large size and long-slim shape with rugged surfaces (low sphericity). Based on the formation mechanism, the gas pores could be further categorized as hydrogen related (small in size) and trapped air related (large in size). Both gas and shrinkage pores can be observed from the correlation graph between pore size and sphericity in Figure 4. This is further discussed in the numerical experiment section (Figure 39). It is also worth noticing that in HPDC process, large pressures during processing tend towards pore sphericities below 0.82. Before proceeding to technical details, for convenience, a summary of critical mathematical symbols used in this work is given in Table 1. Specific meaning should be understood from the context.

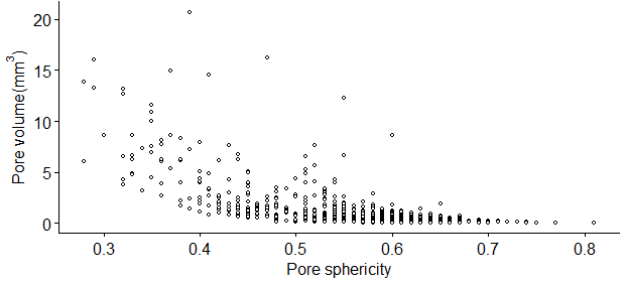

*Figure 4: Correlation between pore volume and pore sphericity.*

*Table 1: Critical mathematical symbols.*

| Symbol | Meaning |
|---|---|
| $\Omega$ | Geometry domain |
| $\Omega_P$ | Domain of pores |
| $\Omega_S$ | Domain with quantity of interests |
| $\Gamma$ | Domain boundary |
| $\Gamma^h$ | Dirichlet boundary |
| $\Gamma^s$ | Neumann boundary |
| $\Gamma^P$ | Pore boundary |
| $\eta$ | Shape parameter |
| $\xi$ | Topological parameter |
| $\Psi$ | Performance function |
| $\Psi_0$ | Performance function on reference domain |
| $n$ | Normal direction |
| $x$ | Point in spatial configuration |
| $X$ | Point in material configuration |
| $z$ | Primary solution |
| $\lambda$ | Adjoint solution |
| $g$ | Generic function |
| $\sigma$ | Stress |
| $\varepsilon$ | Strain |
| $C$ | Material tangent moduli |
| $K$ | Finite element stiffness matrix |
| $f$ | Primary load |
| $f^b$ | Body force |
| $f^s$ | Surface traction |
| $L$ | Adjoint load |
| $T_{topo}$ | Topology sensitivity |
| $T_{shape}$ | Shape sensitivity |
| $D_{topo}$ | Topology sensitivity-based estimator |
| $D_{shape}$ | Shape sensitivity-based estimator |
| $D_{pore}$ | Porosity sensitivity-based estimator |
| $T$ | Shape transformation |
| $V$ | Design speed |
| $\upsilon$ | Poisson's ratio |
| $I_D$ | Effectivity index based on quantity changes |
| $I_\Psi$ | Effectivity index based on quantity values |
| $J$ | Jacobian matrix of domain transformations |
| $U$ | Strain energy |
| $W$ | External work |
| $\Pi$ | Total potential energy |

## 3. PROBLEM STATEMENT

The problem is stated in a smooth bounded 3D domain [41] with an arbitrarily shaped internal pore as shown in Figure 5. The domain can be represented as a difference between a dense structure $\Omega$ and a pore $\Omega_p$, the original domain with fully sized pore could be denoted by $(\Omega-\Omega_p)$, with a boundary $(\Gamma+\Gamma_p)$.

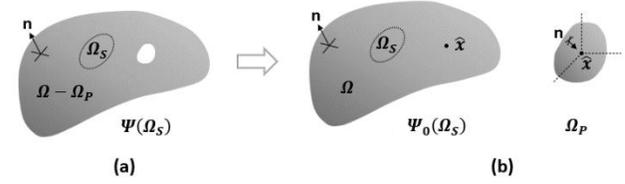

*Figure 5: A flawed geometry with an internal pore in **(a)** could be viewed as the difference between a reference dense geometry and the included pore in **(b)**.*

In this paper, the size of the included pore is assumed much smaller than the structure, i.e.,

$$\|\Omega_P\| < \|\Omega\| \qquad (3.1)$$

and its location is also assumed far away from structure surfaces or other pores, formally

$$\|\Omega_P\| < \text{distance}(\Gamma,\Gamma_P) \qquad (3.2)$$

$$\|\Omega_P\| < \text{distance}(\Gamma_P,\Gamma_P^{other}) \qquad (3.3)$$

It is noted that in this paper the above assumptions are given in order to simplify the proposed model where an arbitrary performance function $\Psi(\Omega_s)$ could be predicted by a first-order design sensitivity-based estimator without considering pore-to-pore or pore-to-surface interactions. However, we will later show that even when the above assumptions are not fully satisfied, for example the W-profile plate with a thin wall structure in the numerical experiment section, the proposed model is still robust, accurate, and can be utilized.

With the above assumptions, suppose a linear elastic boundary value problem [25] could be defined over the porous geometry as

$$\begin{cases} -\nabla \cdot \sigma(z) = f^b & x \in \Omega-\Omega_p \\ z = \hat{z} & x \in \Gamma^h \\ \sigma(z)\cdot n = f^s & x \in \Gamma^s \\ \sigma(z)\cdot n = 0 & x \in \Gamma^P \end{cases} \qquad (3.4)$$

where a prescribed displacement field $z$ is defined over Dirichlet boundary $\Gamma^h$, a surface traction $f^s$ is prescribed on Neumann boundary $\Gamma^s$ with an outward unit normal vector $n$ and a zero Neumann is assumed on pore surfaces $\Gamma^P$.

A generic quantity of interest in a desired region $\Omega_s$ could be calculated as

$$\Psi = \iiint_{\Omega_s} g(z)d\Omega \qquad \Omega_S \subset \Omega-\Omega_p \qquad (3.5)$$

where $g$ represents an arbitrary function dependent on the displacement field $z$.

To solve for $\Psi(\Omega_s)$ on a flawed structure $(\Omega-\Omega_p)$, directly solving Equation (3.4) using FE methods seems straightforward. But several issues arise. First, when there are a large number of irregularly shaped pores, which is usually the case in real applications, volume meshing often leads to highly distorted or inverted elements. Such ill-shaped elements could deteriorate global structural stiffness matrix in FE analysis, leading to convergence problems.

An alternative approach exploits the difference between the two distinct domains, namely $(\Omega-\Omega_p)$ and $\Omega$, by solving a similar boundary value problem on the reference dense structure

$$\begin{cases} -\nabla \cdot \sigma(z) = f^b & x \in \Omega \\ z = \hat{z} & x \in \Gamma^h \\ \sigma(z) \cdot n = f^s & x \in \Gamma^s \end{cases} \quad (3.6)$$

where the performance function over the same region $\Omega_s$ could be defined by

$$\Psi_0 = \iiint_{\Omega_S} g(z) d\Omega \qquad \Omega_S \subset \Omega \quad (3.7)$$

The objective now is to estimate the generic quantity of interests $\Psi(\Omega_s)$ on a porous domain using the performance function defined over the same region ($\Omega_s$) but on a dense structure.

An overview of the proposed method is illustrated in Figure 6 where the transformation of an included pore is highlighted. In Figure 6, let the pore be parameterized by a continuous geometry parameter η on the interval [0, 1]. While the state (η=1) presents the full-sized pore as in Figure 6(a), the range (0<η<1) accounts for a continuously size-shrinking process as in Figure 6(b). In this process, the first order shape sensitivity is employed to account for the change of performance functions with respect to the change of pore shape. At the stage when (η=ξ, ξ≈0⁺), we assume the pore is scaled to an infinitesimally small size (ξ) such that the influence of its original shape could be assumed trivial on structural performances in Figure 6(c). If the pore shape could be assumed as a sphere with equivalent radius ξ, removal of such sphere results in the dense domain in Figure 6(d), where topological sensitivity is utilized to capture the impact of the topological change. Novotny et al. [41] discussed the relationship between the topological parameter and shape parameter in details.

By exploiting shape sensitivity and topological sensitivity, the desired performance function $\Psi(\Omega_s)$ on the porous structure in Figure 6(a) therefore could be approximated by solutions from the dense structure in Figure 6(d) as

$$\Psi(\Omega_S) \approx \Psi_0(\Omega_S) + \mathcal{D}_{topo} + \mathcal{D}_{shape} \quad (3.8)$$

where $\mathcal{D}_{topo}$ and $\mathcal{D}_{shape}$ are estimations from topological sensitivity and shape sensitivity, respectively. By combining the two estimations, we propose a porosity-oriented estimator $\mathcal{D}_{pore}$ as

$$\mathcal{D}_{pore} = \mathcal{D}_{topo} + \mathcal{D}_{shape} \quad (3.9)$$

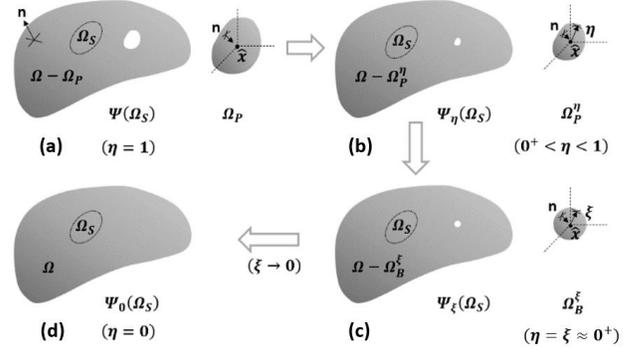

*Figure 6: Overview of the proposed method.*

In the following section, technical backgrounds upon which the proposed methodology is built are reviewed first.

## 4. TECHNICAL BACKGROUND

The proposed porosity sensitivity-based estimator is a combination of two design sensitivity fields, namely, topological sensitivity and shape sensitivity. The concepts of the two design sensitivities are reviewed in this section.

### 4.1 Topological Sensitivity

Topological sensitivity captures the first order impact of inserting an infinitesimally small spherical hole within a domain on various quantities of interests. This concept has its roots in the influential paper by Eschenauer [42], and has later been extended by numerous authors [43]–[48].

Consider the problem illustrated in Figure 7. Let a quantity of interest $\Psi_0(\Omega_s)$ defined within a region of interests $\Omega_s$ on a smooth bounded domain $\Omega$. Suppose an infinitesimal hole with radius ξ is introduced by perturbing the domain $\Omega$ at an arbitrary location $\hat{x}$. A new domain is generated ($\Omega - \Omega_B^\xi$), with the boundary ($\Gamma + \Gamma_B^\xi$). Along with domain topological change, the performance function defined in the same region $\Omega_s$ but on the perturbed domain could be written as [41]

$$\Psi_\xi(\Omega_S) = \Psi_0(\Omega_S) + f(\xi)\mathcal{T}_{topo}(\hat{x}) + R(f(\xi)) \quad (4.1)$$

where $f(\xi)$ is a monotone function whose value tends to zero as (ξ->0). $\mathcal{T}_{topo}(\hat{x})$ is the first-order topological derivative defined at location $\hat{x}$, and $R(f(\xi))$ contains all higher order terms.

If we drop the higher order terms from Equation (4.1), it could be rewritten as the classic topological sensitivity as

$$\mathcal{T}_{topo}(\hat{x}) \equiv \lim_{\xi \to 0} \frac{\Psi_\xi(\Omega_S) - \Psi_0(\Omega_S)}{f(\xi)} \quad (4.2)$$

where the monotone function $f(\xi)$ is taken as the volume of the small spherical hole [41].

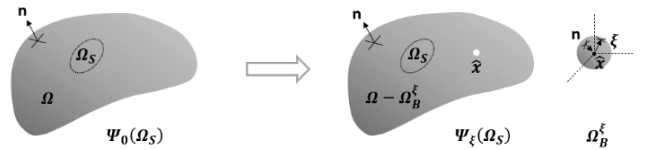

*Figure 7: Illustration of a topological change.*

To find a closed-form expression for the topological sensitivity, one must first define an adjoint. Recall that the adjoint field associated with a quantity of interest satisfies [29], [49]:

$$K\lambda = -\nabla_z \Psi \quad (4.3)$$

where $K$ is the global stiffness matrix in a linear FE system and $\lambda$ is the adjoint solution. The right-hand side of Equation (4.3), i.e., adjoint loads, could be symbolically determined by distribution theory [50] (for details please see Appendix C).

Once the adjoint is computed, topological derivative could be computed on any point over the entire reference dense domain $\Omega$, for instance, for compliance [41]

$$\mathcal{T}_{topo}(\hat{x}) = \frac{3}{4}\frac{1-\nu}{7-5\nu}\left[10\sigma(z):\varepsilon(z) - \frac{1-5\nu}{1-2\nu}tr[\sigma(z)]tr[\varepsilon(z)]\right] \quad (4.4)$$

where the adjoint field is identical to primary displacement solutions, i.e., $\lambda=z$, $\sigma(z)$ and $\varepsilon(z)$ are the stress and strain tensors, and $\nu$ is the Poisson's ratio.

To put things in perspective, once we obtain an analytical expression of topological sensitivity, e.g., Equation (4.5), an arbitrary quantity of interests on a flawed structure with an infinitesimal spherical pore could be approximated by neglecting higher order terms from Equation (4.1) as:

$$\Psi(\Omega_S) \approx \Psi_0(\Omega_S) + Vol(\xi)\mathcal{T}_{topo}(\hat{x}) \quad (4.5)$$

where $Vol(\xi)$ is the volume of the included small pore.

Without considering shape perturbations, the topology sensitivity-based estimator is obtained by comparing Equation (4.5) with (3.8) as:

$$\mathcal{D}_{topo} = Vol(\xi)\mathcal{T}_{topo}(\hat{x}) \quad (4.6)$$

### 4.2 Shape Sensitivity

Next, let us consider a smooth bounded porous geometry ($\Omega - \Omega_P^\eta$) [41] with a parameterized pore ($\Omega_P^\eta$) located at an arbitrary point $X_c$. If the pore geometry is perturbed by shrinking an infinitely small amount ($d\eta$), a perturbed geometry can be expressed as ($\Omega - \Omega_P^{\eta-d\eta}$) with its included pore domain $\Omega_P^{\eta-d\eta}$. We therefore have an original domain with a parameterized pore and a newly perturbed domain with a slightly scaled pore, as shown schematically in Figure 8.

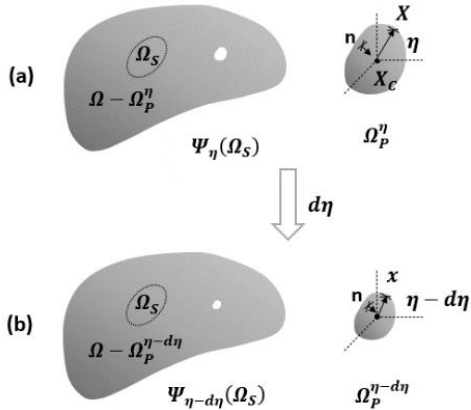

**Figure 8:** Domain transformation on the included pore: **(a)** parameterized geometry, and **(b)** its perturbed counterpart.

If the mapping between the two domains in Figure 8 is assumed smooth and invertible, it could be conveniently described by a domain transformation [29]. Let the point on a parameterized pore surface be $X$ on a material domain, then the same point on the perturbed domain can be described as $x$ on a spatial domain.

If we use a shape parameter $\eta$ to denote the amount of geometry change in the direction of perturbation, then the shape transformation $T$ between the two domains in Figure 8 could be expressed by a linear and continuous mapping as [29]:

$$T(\eta):\left[\Omega - \Omega_P^\eta\right](X) \to \left[\Omega - \Omega_P^{\eta-d\eta}\right](x) \quad (4.7)$$

For a point on the pore surface, the mapping represents a scaling process where the pore is reduced to a slightly smaller size:

$$T(\eta): x = (X - X_c)\eta + X_c \qquad x \in \Gamma_p \quad (4.8)$$

For any other point *not* on the pore surface, the mapping would not change its location:

$$T: x = X \qquad x \notin \Gamma_P \quad (4.9)$$

If the perturbation process is considered to relate pseudo-time, a design speed $V$ can be defined as [29]

$$V \equiv \frac{dT}{d\eta} \quad (4.10)$$

Specifically, the design speed on pore surfaces $\mathbf{x} \in \Gamma_p$ could be computed as

$$V = X - X_c \quad (4.11)$$

On the other hand, the design speed, which is *not* on pore boundaries, is assumed as zero

$$V = 0 \qquad \forall x \notin \Gamma_P \quad (4.12)$$

Based on the predefined design speeds, shape sensitivity for any generic quantity of interest in region $\Omega_s$ could be derived through material derivative in either domain or boundary forms [29] (for details please see Appendix C and D).

The boundary form is preferred in this work for several reasons. First, with tomography reconstruction as an enabling technology, it provides detailed geometry description on pore surfaces. The surfaces can be studied in the transformation in Figure 8, which also gives an explicit expression for design speed. Second, compared with 3D FE volume mesh, mesh generation and computation on a surface mesh is less error prone and computationally efficient.

A uniform equation of shape sensitivity for a linear elastic performance can be derived in the form of porosity boundary integral as (for derivations please see Appendix C and D)

$$\frac{d(\Psi_\eta(\Omega_s))}{d\eta} = -\iint_{\Gamma_\eta^P}\left[\sigma(z_\eta):\varepsilon(\lambda_\eta)\right]V_n^\eta d\Gamma \quad (4.13)$$

where $\Gamma_\eta^P$ refers to the parameterized pore boundaries with normal design speeds $V_n^\eta$, $\sigma(z_\eta)$ and $\varepsilon(\lambda_\eta)$ are the stress and strain fields on pore surfaces computed from primary and adjoint fields, respectively. Since design speed is only defined on the boundary of pores, the boundary integration therefore only needs to be carried out on the pore surfaces.

It is noted that evaluation of Equation (4.13) requires both the primary displacement fields and adjoint solutions. While the primary solutions can be obtained by solving Equation (3.6), the adjoint solutions are achievable by the Equations (C.12), (C.19), and (C.25). It is also noted that solving for adjoints would be trivial if the stiffness matrix associated with the primary problem is already factored, since the only difference between the two systems is the load definition which appears on the right-hand side of a classic FE equation.

## 5. PROPOSED METHOD

Given the topological and sensitivity fields, in this section, a porosity sensitivity-based estimator is proposed to predict an arbitrary linear elastic performance function on porous structures with internal pores. The proposed method starts with solving shape parameter (η) dependent problems.

### 5.1 Parameter dependent formulations

Let us first use the pore shape parameter η to generalize the boundary value problem in Equation (3.4).

Then, for a structure with a fully sized pore (η=1):

$$\begin{cases} -\nabla \cdot \sigma(z) = f^b & x \in \Omega - \Omega_p \\ z = \hat{z} & x \in \Gamma^h \\ \sigma(z) \cdot n = f^s & x \in \Gamma^s \\ \sigma(z) \cdot n = 0 & x \in \Gamma^P \end{cases} \quad (5.1)$$

For a porous structure with a parameterized pore (0<η<1):

$$\begin{cases} -\nabla \cdot \sigma(z_\eta) = f^b & x \in \Omega - \Omega_P^\eta \\ z_\eta = \hat{z} & x \in \Gamma^h \\ \sigma(z_\eta) \cdot n = f^s & x \in \Gamma^s \\ \sigma(z_\eta) \cdot n = 0 & x \in \Gamma_P^\eta \end{cases} \quad (5.2)$$

For a dense structure without pores (η=0):

$$\begin{cases} -\nabla \cdot \sigma(z) = f^b & x \in \Omega \\ z = \hat{z} & x \in \Gamma^h \\ \sigma(z) \cdot n = f^s & x \in \Gamma^s \end{cases} \quad (5.3)$$

It is noted at the state (η=1) the parameterized boundary value problem in Equation (5.2) resembles Equation (5.1), and when (η=0), the parameterized formulation reduces to Equation (5.3).

Similarly, the performance function defined in Equation (3.5) could be also parameterized.

For a structure with a full-size pore (η=1):

$$\Psi = \iiint_{\Omega_S} g(z) d\Omega \qquad \Omega_S \subset \Omega - \Omega_p \quad (5.4)$$

For a porous structure with a parameterized pore (0<η<1):

$$\Psi_\eta = \iiint_{\Omega_S} g(z_\eta) d\Omega \qquad \Omega_S \subset \Omega - \Omega_P^\eta \quad (5.5)$$

For a dense structure without pores (η=0):

$$\Psi_0 = \iiint_{\Omega_S} g(z) d\Omega \qquad \Omega_S \subset \Omega \quad (5.6)$$

In a similar manner, when (η=1), the parameterized quantity of interests in Equation (5.5) is identical to that on a fully sized porous structure in Equation (5.4). When (η=0), parameterized performance function is then reduced to Equation (5.6).

If the performance function $\Psi_\eta$ is assumed to be a continuously differentiable function, then based on the fundamental theorem of calculus, the desired performance measure Ψ on a fully flawed structure can be calculated through

$$\Psi = \Psi_0 + \int_0^1 \left( \frac{d\Psi_\eta}{d\eta} \right) d\eta \quad (5.7)$$

As discussed in Section 3, the range of porosity scaling parameter η could be divided into two: a size shrinking process (ξ≤η≤1) and an infinitesimally small hole removal process (0≤η≤ξ) with (ξ=0⁺). That is, the last term in Equation (5.7) could be expanded according to the two processes as:

$$\int_0^1 \left( \frac{d\Psi_\eta}{d\eta} \right) d\eta = \int_0^\xi \left( \frac{d\Psi_\eta}{d\eta} \right) d\eta + \int_\xi^1 \left( \frac{d\Psi_\eta}{d\eta} \right) d\eta \quad (5.8)$$

By combining with Equation (5.7), we have

$$\Psi = \Psi_0 + \int_0^\xi \left( \frac{d\Psi_\eta}{d\eta} \right) d\eta + \int_\xi^1 \left( \frac{d\Psi_\eta}{d\eta} \right) d\eta \quad (5.9)$$

By comparing with Equation (3.8), it could be readily seen

$$\mathcal{D}_{topo} = \int_0^\xi \left( \frac{d\Psi_\eta}{d\eta} \right) d\eta \quad (5.10)$$

$$\mathcal{D}_{shape} = \int_\xi^1 \left( \frac{d\Psi_\eta}{d\eta} \right) d\eta \quad (5.11)$$

On invoking Equation (4.6), the Equation (5.10) could be approximated by a topological sensitivity field as

$$\mathcal{D}_{topo} = \int_0^\xi \left( \frac{d\Psi_\eta}{d\eta} \right) d\eta \approx Vol(\xi) \mathcal{T}_{topo}(\hat{x}) \quad (5.12)$$

On the other hand, by invoking Equation (4.13), accounting for the integrand in Equation (5.11) needs solving the following adjoint equations:

$$\begin{cases} -\nabla \cdot \sigma(\lambda_\eta) = L_b^\eta & x \in \Omega_\eta \\ \lambda_\eta = 0 & x \in \Gamma_\eta^h \\ \sigma(\lambda_\eta) \cdot n = L_s^\eta & x \in \Gamma_\eta^s \end{cases} \quad (5.13)$$

where $\lambda_\eta$ is the adjoint solution, $L_b^\eta$ and $L_s^\eta$ are adjoint body force and adjoint surface traction, respectively. The adjoint load definitions could be different for different performance functions. As for the derivations of adjoints for various linear elastic performances studied in this paper (elastic compliance, local averaged displacement, and nodal displacement) please refer to Appendix C and D.

By comparing Equation (5.11) and (4.13), it can be observed that the integration of Equation (4.13) is a function of the porosity scaling parameter η. Computing the integration in Equation (5.11) requires evaluation of the shape sensitivity in Equation (4.13) for all η on [ξ, 1], which is impractical.

Therefore, in the following sections, we approximate all the fields in Equation (4.13), i.e., $\sigma(z_\eta)$, $\varepsilon(\lambda_\eta)$, $V_n^\eta$ and $\Gamma_P^\eta$, as explicit functions of η. Then, through analytical integration, evaluation of Equation (5.11) become feasible.

## 5.2 Exterior approximation

The objective of this section is to use exterior formulation [34] to approximate field variables on pore surfaces as an explicit function of shape parameter η. For a linear elastic problem, its boundary value problem could be written for a reference dense domain as

$$\begin{cases} -\nabla \cdot \sigma(z_0) = f_b & x \in \Omega \\ z_0 = \hat{z} & x \in \Gamma^h \\ \sigma(z_0) \cdot n = f_s & x \in \Gamma^s \end{cases} \quad (5.14)$$

The same governing equations but on a porous structure with a parameterized pore can be written as

$$\begin{cases} -\nabla \cdot \sigma(z_\eta) = f_b & x \in \Omega - \Omega_p^\eta \\ z_\eta = \hat{z} & x \in \Gamma^h \\ \sigma(z_\eta) \cdot n = f_s & x \in \Gamma^s \\ \sigma(z_\eta) \cdot n = 0 & x \in \Gamma_p^\eta \end{cases} \quad (5.15)$$

We assume there could be a linear relation between the two displacement fields as

$$z_\eta = z_0 + \tilde{z}_\eta \quad (5.16)$$

From constitutive equations of linear elastic materials,

$$\begin{cases} \sigma_0 = C\varepsilon_0 = \frac{1}{2}C(\nabla z_0 + \nabla z_0^T) \\ \sigma_\eta = C\varepsilon_\eta = \frac{1}{2}C(\nabla z_\eta + \nabla z_\eta^T) \end{cases} \quad (5.17)$$

it is ready to see that

$$\sigma(\tilde{z}_\eta) = \sigma(z_\eta - z_0) = \sigma(z_\eta) - \sigma(z_0) \quad (5.18)$$

Combining Equation (5.14) with (5.15) and (5.18), it could be shown

$$\begin{cases} -\nabla \cdot \sigma(\tilde{z}_\eta) = 0 & x \in \Omega - \Omega_p^\eta \\ \tilde{z}_\eta = 0 & x \in \Gamma^h \\ \sigma(\tilde{z}_\eta) \cdot n = 0 & x \in \Gamma^s \\ \sigma(\tilde{z}_\eta) \cdot n = -\sigma_0 \cdot n & x \in \Gamma_p^\eta \end{cases} \quad (5.19)$$

Recall the assumptions in Equation (3.1) and (3.2). That is, if a pore is much smaller than the structure and it is located far away from structure surface, the residual field $\tilde{z}_\eta$ in Equation (5.19) could be approximated by $\tilde{z}_\eta^*$ in an exterior Neumann condition as [51]:

$$\begin{cases} -\nabla \cdot \sigma(\tilde{z}_\eta^*) = 0 & x \in R^n - \Omega_p^\eta \\ \tilde{z}_\eta^* = 0 & x \to \infty \\ \sigma(\tilde{z}_\eta^*) \cdot n = -\sigma_0 \cdot n & x \in \Gamma_p^\eta \end{cases} \quad (5.20)$$

The exterior approximation could be solved either through boundary element methods [52] or FE methods with infinite elements [53].

Now consider the displacement field $z=z(x; \eta)$ defined over a parameterized spatial domain $\Omega_\eta$. Its material representation which is defined on the original (material) domain $\Omega_0$ could be obtained by exploiting the transformation $x= x(X; \eta)$.

By assuming the displacement fields over the two domains are identical,

$$z(x;\eta) = z(X;\eta) \quad (5.21)$$

wherein the coordinates in the two domains could be mapped through Equation (4.8).

Then any point on the parameterized pore surface $\Gamma_P^\eta$ could therefore be written in the form of material domain

$$\Gamma_P^\eta = \eta \Gamma_P \quad (5.22)$$

where $\Gamma_P$ refers to the fully scaled pore surface (η=1).

Recall the design speed defined for the transformation on a parameterized pore surface in Equation (4.11). It can be expressed in the material domain as

$$V_n^\eta = V_n \quad (5.23)$$

If we take gradients of the displacement field with respect to coordinates of the two domains,

$$\nabla z(x) = J \nabla Z(X) \quad (5.24)$$

where $J$ is the Jacobian matrix between the two domains.

Combining Equation (5.17) and (5.24), it could be shown the relation of stress fields during the domain transformation as:

$$\sigma(x) = \frac{1}{\eta}\sigma(X) \quad (5.25)$$

Substitute Equation (5.25) into (5.20):

$$\begin{cases} -\nabla \cdot \sigma(\tilde{z}_\eta^*(X)) = 0 & X \in R^n - \Omega_p^1 \\ \tilde{z}_\eta^*(X) = 0 & X \to \infty \\ \frac{1}{\eta}\sigma(\tilde{z}_\eta^*(X)) \cdot n = -\sigma_0 \cdot n & X \in \Gamma_P^1 \end{cases} \quad (5.26)$$

and let

$$z_E = \frac{1}{\eta}\tilde{z}_\eta^*(X), \ \sigma(z_E) = \frac{1}{\eta}\sigma(\tilde{z}_\eta^*(X)) \quad (5.27)$$

plug into Equation (5.26), the exterior approximation could be formulated *independent* of the shape parameter $\eta$ as:

$$\begin{cases} -\nabla \cdot \sigma(z_E) = 0 & X \in R^n - \Omega_p^1 \\ z_E = 0 & X \to \infty \\ \sigma(z_E) \cdot n = -\sigma_0 \cdot n & X \in \Gamma_P^1 \end{cases} \quad (5.28)$$

The stress and strain fields can be written in an *explicit* form of the shape parameter $\eta$ as:

$$\begin{aligned} \sigma(z_\eta(x)) &\approx \sigma(z_0) + \sigma(\tilde{z}_\eta^*(x)) \\ &= \sigma(z_0) + \frac{1}{\eta}\eta\sigma(z_E(X)) \\ &= \sigma(z_0) + \sigma(z_E(X)) \end{aligned} \quad (5.29)$$

$$\varepsilon(z_\eta(x)) \approx \varepsilon(z_0) + \varepsilon(z_E(X)) \tag{5.30}$$

In a similar manner, the exterior approximation can be defined on *adjoint* fields as

$$\begin{cases} -\nabla \cdot \sigma(\lambda_E) = 0 & X \in R^n - \Omega_p^1 \\ \lambda_E = 0 & X \to \infty \\ \sigma(\lambda_E) \cdot n = -\sigma_0 \cdot n & X \in \Gamma_p^1 \end{cases} \tag{5.31}$$

whereas its stress and strain fields can be written as

$$\sigma(\lambda_\eta(x)) \approx \sigma(\lambda_0) + \sigma(\lambda_E(X)) \tag{5.32}$$

$$\varepsilon(\lambda_\eta(x)) \approx \varepsilon(\lambda_0) + \varepsilon(\lambda_E(X)) \tag{5.33}$$

It is noted if the stress or strain fields, which are on the dense structure and close to pore surfaces, do not exhibit significant fluctuation, the fields can be further simplified to the value at the pore center from Equation (5.29) to (5.33).

## 5.3 Porosity sensitivity

Since all the terms ($\sigma(z_\eta)$, $\varepsilon(\lambda_\eta)$, $V_n^\eta$ and $\Gamma_p^\eta$) in Equation (4.13) could be expressed as explicit functions of η through exterior approximation, the integration of shape sensitivity in Equation (5.11) could therefore be calculated:

$$\begin{aligned} \mathcal{D}_{shape} &= \int_\xi^1 \mathcal{T}_{shape} d\eta \\ &\approx \int_\xi^1 \left( \int_{\Gamma_p^\eta} (-\sigma_\eta(z) : \varepsilon_\eta(\lambda)) V_n^\eta d\Gamma_\eta \right) d\eta \\ &= \int_{\Gamma^p} \left\{ \begin{array}{l} \int_\xi^1 -[\sigma_0(z) + \sigma_E(z)] : \\ [\varepsilon_0(\lambda) + \varepsilon_E(\lambda)] \eta V_n d\eta \end{array} \right\} d\Gamma \\ &= \int_{\Gamma^p} \left\{ \begin{array}{l} -\dfrac{V_n}{2} [\sigma_0(z) + \sigma_E(z)] : \\ [\varepsilon_0(\lambda) + \varepsilon_E(\lambda)](1 - \xi^2) \end{array} \right\} d\Gamma \end{aligned} \tag{5.34}$$

Therefore, the porosity sensitivity-based estimator could be calculated by summing contributions from topological and shape sensitivities as

$$\mathcal{D}_{pore} = \mathcal{D}_{topo} + \mathcal{D}_{shape} \tag{5.35}$$

where $\mathcal{D}_{topo}$ is defined in Equation (5.12) and $\mathcal{D}_{shape}$ is computed from Equation (5.34).

Therefore, an arbitrary linear elastic quantity of interests could be approximated through porosity sensitivity-based estimator as

$$\Psi(\Omega_S) = \Psi_0(\Omega_S) + \mathcal{D}_{pore} \tag{5.36}$$

## 5.4 Multiple pores

In a typical manufactured metal component, the number of included pores is often many. Therefore, the proposed porosity sensitivity-based estimator for a single pore in Equation (5.35) has to be extended to scenarios with multiple pores.

For estimating arbitrary quantity of interests, if we consider each pore as a design variable and the proposed porosity sensitivity as a first-order derivative, then the performance function on a porous structure could be estimated by

$$\Psi = \Psi_0 + \sum_{i=1}^N D_{pore}^i \tag{5.37}$$

where higher order terms are neglected, and $D_{pore}^i$ is the porosity sensitivity-based estimator for the i[th] pore with a total of N pores.

In other words, the behavior of a porous structure could be estimated by simply adding up independently computed $D_{pore}^i$ on each pore. Inclusion of higher order pore-to-pore or pore-to-surface interaction terms was found more precise but also more computationally expensive [32].

## 5.5 Proposed Algorithm

The overall algorithm is illustrated in Figure 9 and described below.

1. With manufactured flawed component (Ω) tomography scanned, the geometry of its internal pores ($\Omega_p$) is reconstructed from scanning images. The geometry of the (dense) component without pores (Ω-$\Omega_p$) can be obtained either from CT-reconstruction or directly from a CAD model.

2. Impose proper loading and boundary conditions on the dense structure (Ω-$\Omega_p$) where primary solutions (*z*) are obtained by solving Equation (3.6) and adjoint fields (λ) are computed through Equation (C.12), (C.19), and (C.25). With resulting primary and adjoint solutions, topological sensitivity is computed on each material point on the dense structure, for example by Equation (4.4), while the topological sensitivity-based estimator ($D_{topo}$) is computed via Equation (4.6). Quantities of interests ($\Psi_0$) on the dense structure is calculated by Equation (5.6).

3. For each pore ($\Omega_p$), its stress field is approximated by exterior formulation in Equation (5.32) based on the primary and adjoint solutions obtained from Step-2 at the pore location. Its design speed on boundary is computed by Equation (4.11) and (5.23). With the design speed ($V_n$) and approximated porosity stress, the shape sensitivity with respect to quantities of interests could be computed by Equation (4.13). The shape sensitivity-based estimator ($D_{shape}$) is then computed by Equation (5.34).

4. The proposed porosity sensitivity-based estimator is assembled via Equation (5.35).

5. With the initial value of quantity of interests ($\Psi_0$) on dense structure, the quantity value (Ψ) on the flawed structure is predicted by Equation (5.36) and (5.37).

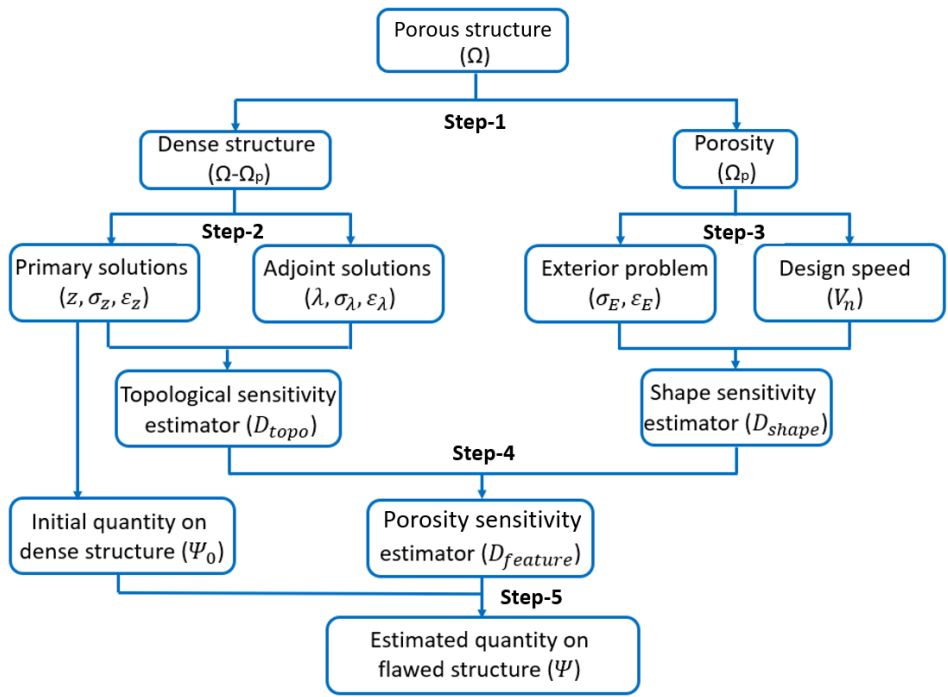

*Figure 9: Proposed algorithm.*

## 6. NUMERICAL EXPERIMENTS

In this Section, the proposed method is demonstrated through several numerical experiments. The proposed porosity sensitivity-based estimator has been implemented in PYTHON for both 2D and 3D cases. The linear elastic problems in Equation (3.4) and (3.6) are solved in ABAQUS [54], a commercial FE package, where the adjoint loads, defined in Equation (C.12), (C.19) and (C.25), are implemented in a user defined subroutine DLOAD. The exterior approximation in Equation (5.28) is solved by Fast-BEM [52], a boundary element method package. Statistical analysis of tomography scanning data has been developed on RSTUDIO [55]. Other software has also been utilized for image processing, geometry repair and mesh editing. Their applications are discussed in the sections where they are applied.

All experiments were conducted on a 64-bit Windows 10 machine with the following hardware: Intel I5-8250U CPU 4 cores running at 1.6GHz with 16 GB of installed physical memory (RAM).

In all experiments, we neglect any metal polycrystalline microstructures (e.g., grain boundary, triple or quad points) except for pores and assume materials are isotropic with perfect linear elastic properties (E=6.89e10 N/m$^2$, v=0.35). Material yield, nonlinearity, plasticity and deformation analysis are beyond the scope of this work. The first example involves a 2D benchmark study in which various pore parameters are studied for their influence on the accuracy of the proposed method. The second example is a case study on a 2D control arm with a simulated porosity distribution. The third example involves a 3D control arm with less than 100 pores where pore surface geometries are regenerated by tomography scanning data. The last example is on a real industrial cast component with more than 1300 pores. Statistical distributions of pore parameters are studied, and 3D models are developed for pore geometries. Quantities of interests are computed in each of the examples, and the proposed porosity sensitivity-based estimator is then compared with two other estimators as well as solution from a direct FE analysis:

- Direct FE analysis: a flawed model with internal pores is created at first hand. The model is then solved in ABAQUS. The quantity of interests is denoted as $\Psi$, which is considered as the ground truth when comparing with estimators. This approach is computational expensive and hence should be avoided in practice.

- Topological sensitivity: field variables (i.e., displacements) are first solved on a dense structure without any pore where its quantity of interests is denoted as $\Psi_o$. If each pore could be represented by an equivalent circle hole in 2D (or a sphere in 3D) at the same location, the estimation on the change of quantity could be computed by multiplying topological sensitivity with the pore's equivalent area (or volume in 3D), denoted as $D_{topo}$ as in Equation (4.6) (please see Section 4). As shown in experiments, this approach is inaccurate and only used here for comparison purpose.

- Shape sensitivity: the estimation on the change of quantity $D_{shape}$ (please see Section 5) could be computed in Equation (5.34) which takes real pore shapes into account but neglect the effect of initial topological change.

- Porosity sensitivity: by combining topological and shape sensitivities, the change of quantity could be

computed though Equation (5.35) and is denoted as $D_{pore}$. This estimator is proved superior to the other two estimators in parameter studies and then applied to case studies.

In order to quantify the accuracy of different estimators, an effectivity index could be defined as the ratio between the estimated value of quantity of interest and its exact value, or the ratio between the predicted change of quantity and exact value [31]. That is,

$$I_\Psi = \frac{\Psi_0 + D}{\Psi} \quad (6.1)$$

$$I_D = \frac{D}{\Psi - \Psi_0} \quad (6.2)$$

where $D$ could be $D_{topo}$, $D_{shape}$, or $D_{pore}$. It is obvious that the closer to 1.0, the more accurate an estimator is. In the field of error estimation for quantities measured in global norm (e.g., compliance), if the effectivity indices are in the range between 0.5 and 2.0, it would be considered as acceptable [31]. However, for a local quantity estimation, e.g., local stress or displacement, the effectivity index for a good estimator could be relaxed up to 10.0, since it is generally much more difficult and expensive to obtain [56].

## 6.1 Benchmark study

In this section, a 2D cantilever beam is used as a benchmark example to study the influences of various pore parameters on the estimation of performance functions.

Pore size

As shown in Figure 10, the dimension of the beam is 200 mm × 100 mm. It is clamped at left side and a uniform pressure (F=1000 N/mm$^2$) is applied on the top surface. The quantity of interests includes elastic compliance, the averaged vertical displacement in a region ($\Omega_s$), and a vertical displacement at a selected node ($P_t$). A circular hole is inserted in the middle of the beam where its radius is varied. The size of the hole varies from 0.5 mm to 20 mm in this study. It is noted the hole size does not indicate a real pore size in metal parts. It is only used here for parameter study.

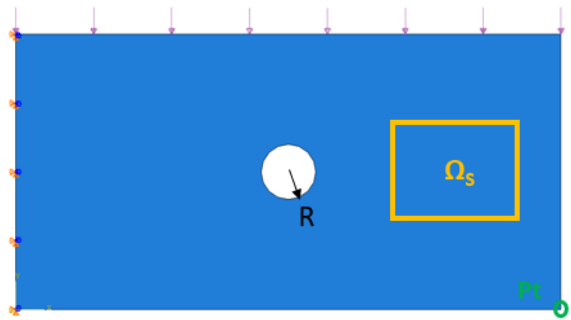

*Figure 10: Parameter study of pore size on a simple 2D cantilever beam where boundary conditions and regions of interests are prescribed.*

In order to show the accuracy of different estimators, the performance functions predicted by topological sensitivity, shape sensitivity and porosity sensitivity are compared from Figure 11 to Figure 13. In these figures, FEA is solved using ABAQUS on the porous domain; the resulting global compliance, nodal displacement and region-averaged displacements are used as ground truth when comparing against the proposed estimators. Three different estimators are computed (topological sensitivity estimator by Equation (4.6), shape sensitivity by Equation (5.34) and the proposed porosity sensitivity by Equation (5.35)) on a dense structure to approximate the performance functions when the circular pore is inserted in the middle of the domain as in Figure 10. It is noted that the closer to FEA results, the more accurate an estimator is.

Even though the estimations from topological sensitivities are in good agreement with FEA ground truth, it is clear that estimations from shape and porosity sensitivities are much more accurate. In fact, the estimations from shape sensitivity and porosity sensitivity are close to each other as illustrated in Figure 11 to Figure 13, and their difference could be observed in Figure 14 through Figure 16.

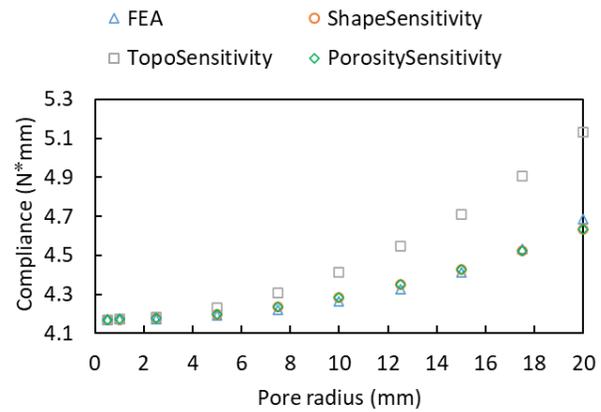

*Figure 11: Comparison of different estimators on the accuracy of compliance.*

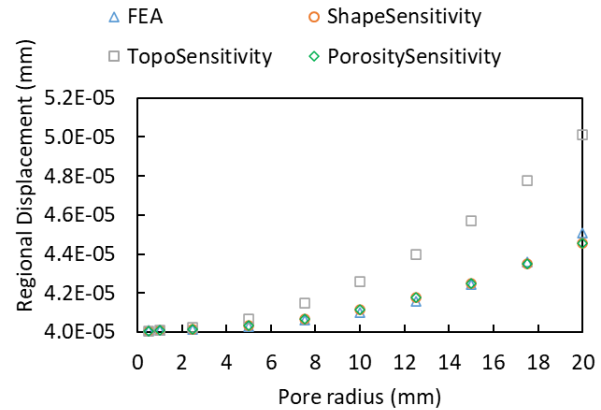

*Figure 12: Comparison of different estimators on the accuracy of regional averaged displacement.*

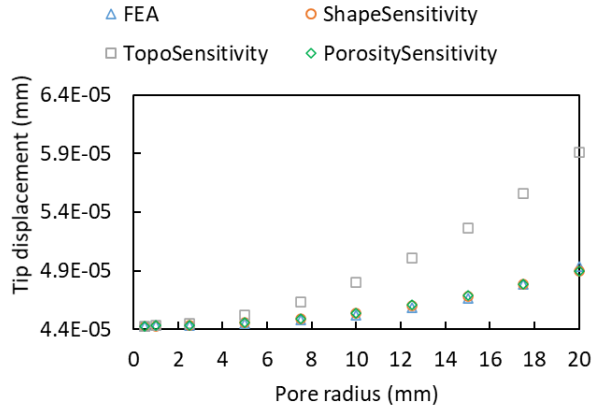

*Figure 13: Comparison of different estimators on the accuracy of nodal displacement.*

In order to further distinguish the three estimators, their effectivity indices in Equation (6.2) are plotted from Figure 14 to Figure 16 where in the figure legend TipDisp represents a vertical displacement at a selected node ($P_t$) in Figure 10 and RegDisp refers to the averaged vertical displacement in the region ($\Omega_s$). By comparing topological and shape sensitivities, it is observed topological sensitivity is more accurate for small pores (pore radius $\leq$ 1mm in this study) while shape sensitivity gives more accurate predictions on large pores. *Porosity sensitivity combines the advantages of the two estimators in that the predictions from topological derivative dominates the lower range of pore sizes and shape derivative governs the higher range so that the prediction is closer to the ground truth in a global sense.* The porosity sensitivity-based estimator is therefore applied onto other experiments.

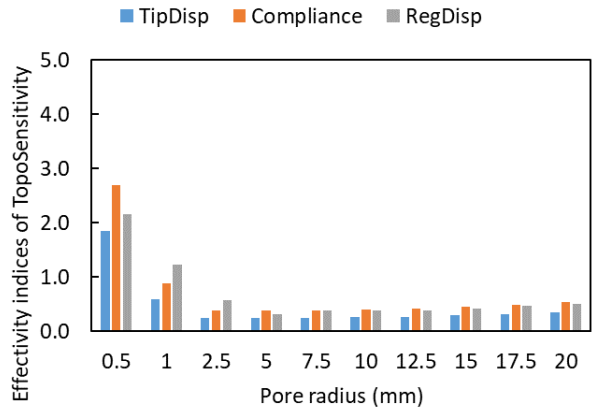

*Figure 14: Effectivity indices of topological sensitivity on different quantities of interests for pore size study.*

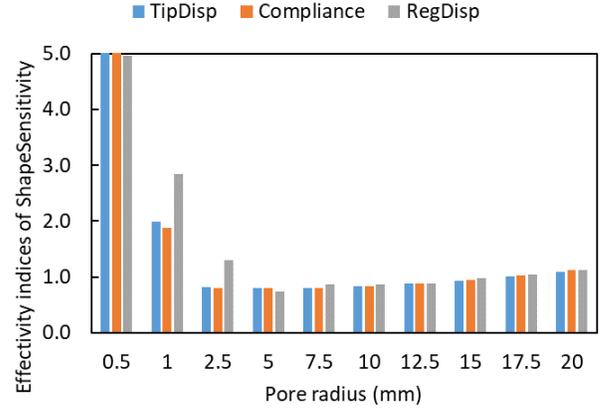

*Figure 15: Effectivity indices of shape sensitivity on different quantities of interests for pore size study.*

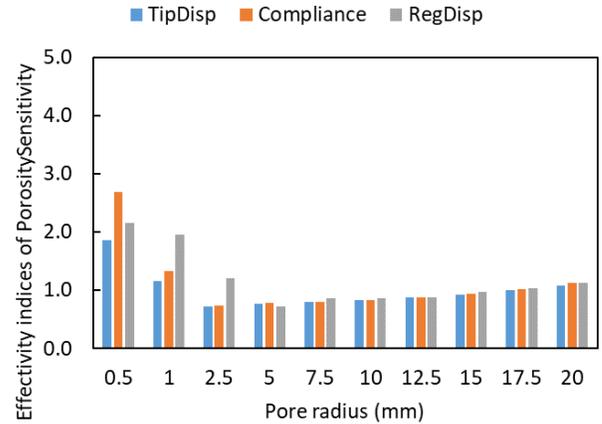

*Figure 16: Effectivity indices of porosity sensitivity on different quantities of interests for pore size study.*

Distance to surface

We have assumed the pore is far away from surface in Equation (3.2). This is because geometry induced stress concentration or singularity could cause high stress gradients near pore surfaces. This high stress gradients could be problematic for exterior formulation in Section 5.2, which assumes the stress fields in the neighborhood of pores almost constant. In this section, we vary the distance between surface and a fix-sized pore to study its influence on the proposed estimator in Figure 17.

It is observed in Figure 18 when the pore is in close distance to surface (e.g., D/2R < 0.2), there could be an estimation error up to 20%. However, if the distance increases to more than 30% of the pore size, the prediction tends to be much more accurate. This observation is therefore consistent with the assumption discussed previously. Therefore, this study shows the further the pore is away from surfaces, the less influence stress gradients would have on approximations, thus more accurate estimations.

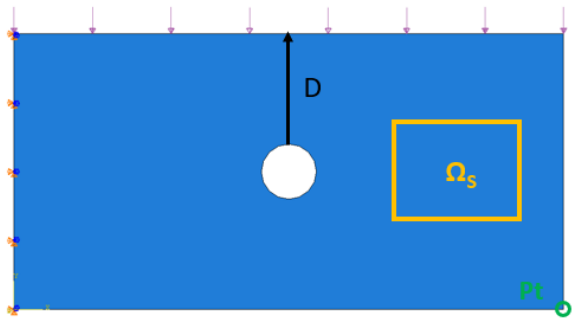

*Figure 17: Parameter study of pore-to-surface distance on a simple 2D cantilever beam where boundary conditions and regions of interests are prescribed.*

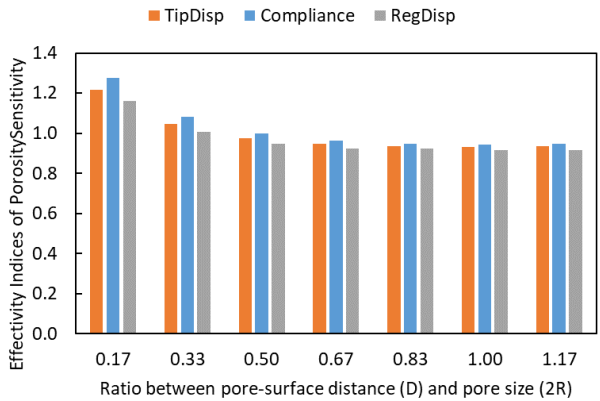

*Figure 18: Effectivity indices of porosity sensitivity on different quantities of interests for pore-to-surface distance study.*

Distance between pores

We have assumed that the distance between different pores is considerably large in Equation (3.3). This is because higher order terms accounting for pore-to-pore interactions are neglected in porosity sensitivity-based estimators. However, in real situations, manufacturing induced pores could be in close distances to others. Therefore, it is necessary to study the influence of distance between pores on the estimation accuracy.

In order to study the interactions between pores, two pores of the same size are placed in the geometry and the distance between them are varied as shown in Figure 19.

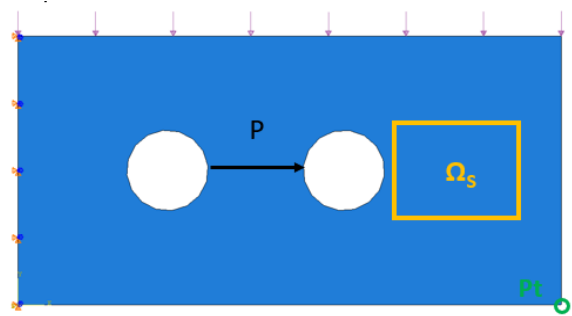

*Figure 19: Parameter study of pore-to-pore distance on a simple 2D cantilever beam where boundary conditions and regions of interests are prescribed.*

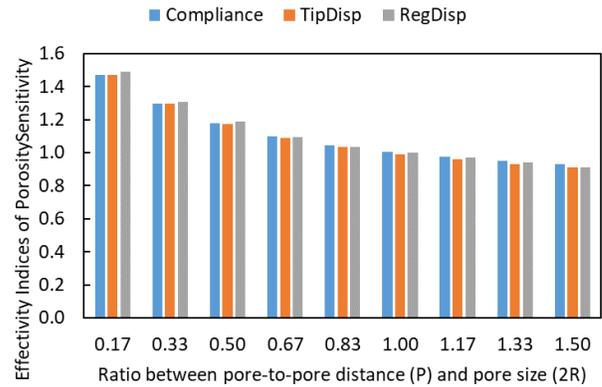

*Figure 20: Effectivity indices of porosity sensitivity on different quantities of interests for pore-to-pore distance study.*

It is observed from Figure 20 that, if the distance is at least the same size of a pore, the porosity sensitivity estimator is highly accurate. On the other hand, if the distance is smaller than 40% of pore size, prediction error (Equation (6.2)) could be up t0 40%, indicating a strong impact of pore-to-pore interactions.

Pore density

From the above pore distance study, we have shown that the pore-to-pore interactions can be a significant factor. In commercial manufactured components, pores are often spatially distributed with different local densities [57]. Pore density, as a measure of the quantity of pores in a locally predefined region, is therefore another important factor worthy of investigation. In this benchmark example, three density levels are studied and the accuracies of their porosity sensitivity estimators are compared.

As shown in Figure 21, three cantilever beams are subject to the same loading and boundary conditions as the previous benchmark examples. Three quantities of interests (global compliance, nodal displacement and regional averaged displacements) are the same as before. On the three beams, we predefine a region ($\Omega_D$) within which different number of identically sized pores are evenly distributed.

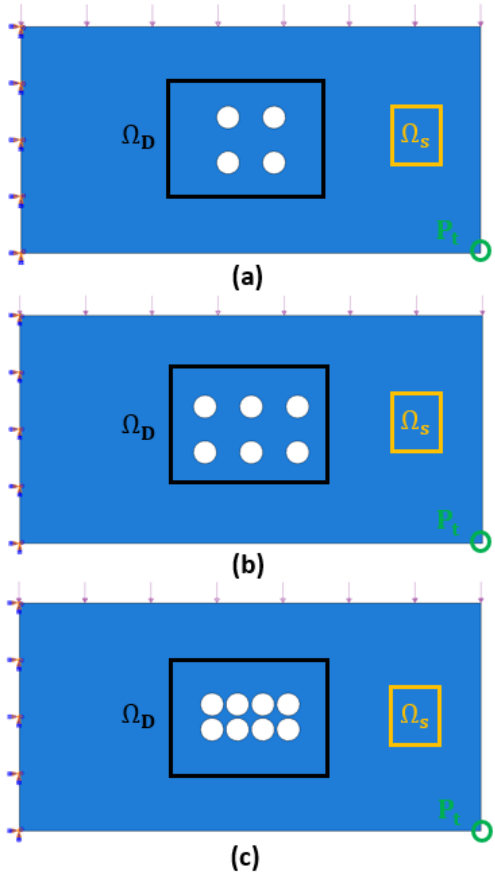

*Figure 21: Parameter study of pore density in a predefined region $\Omega_D$: **(a)** low density, **(b)** medium density, and **(c)** high density*

The influence of pore density on the accuracy of the proposed estimator can be observed in Figure 22. It is evident that with the increase of pore density in Figure 21 (a)–(c), inter-pore distances decrease and pore-to-pore interactions increase. Since higher order interaction terms are neglected in our proposed linear estimator, the stronger interactions in highly densed porous region result in less accurate estimations, consistent with the observations in the pore distance study.

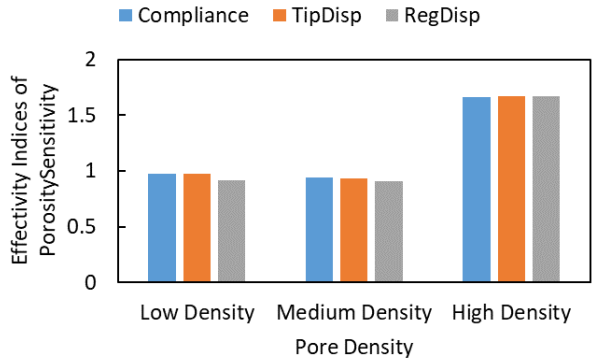

*Figure 22: Effectivity indices of porosity sensitivity estimator on various quantities of interests for pore density study.*

### Pore sphericity

Sphericity is an important parameter to characterize pore shapes. It is defined as the ratio between the surface area of a sphere with the same value as a selected pore and the true surface area of the pore. Various pore shapes could be observed in manufacturing. As discussed in Section 2, a solidification induced pore tends to have a long slim shape with rugged surface while a gas pore is more likely to have a smooth spherical shape. However, pore shapes should not strongly affect the accuracy of the proposed estimator because the influence of pore shapes can be properly accounted for by integrated shape sensitivity as in Equation (3.9).

In order to test the impact of pore shapes on estimations, a single pore with various sphericity values are placed in the middle of the domain whose results are then compared with their FE counterparts, as seen in Figure 23.

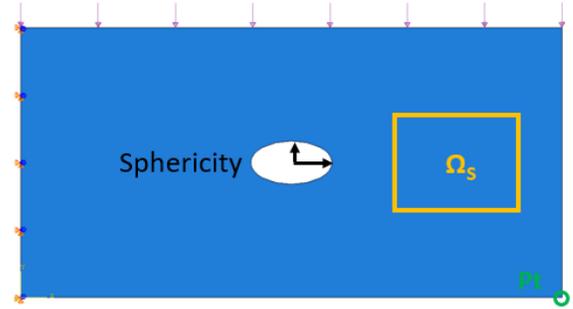

*Figure 23: Parameter study of pore sphericity on a simple 2D cantilever beam where boundary conditions and regions of interests are prescribed.*

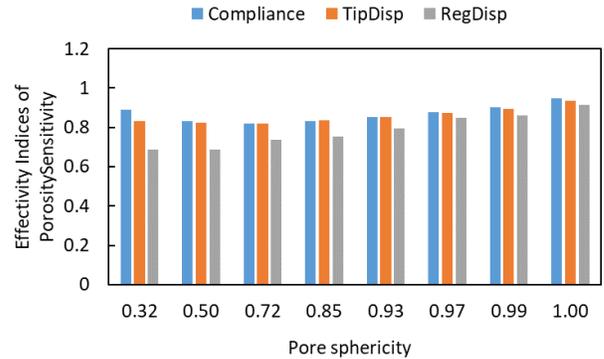

*Figure 24: Effectivity indices of porosity sensitivity on different quantities of interests for pore sphericity study.*

As shown in Figure 24, the sphericity does have a slight influence on estimations: as sphericity decreases to 0.32, the error (Equation (6.2)) increases to around 20%. One plausible reason could be that the stress distribution around the pores of low sphericity could be significantly different than that at its geometrical center which is used in exterior approximation.

### Stress approximation

Stress on pore surfaces plays an essential role in determining the locations of structure cracks or material failures. Therefore,

the accuracy of stress fields approximated by exterior formulation in the proposed model is studied.

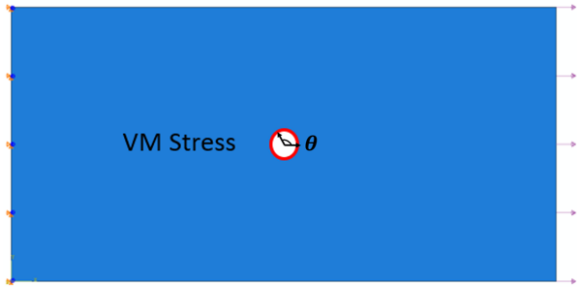

*Figure 25: Stress approximation on pore surfaces (red region) on a simple 2D cantilever beam where boundary conditions are prescribed.*

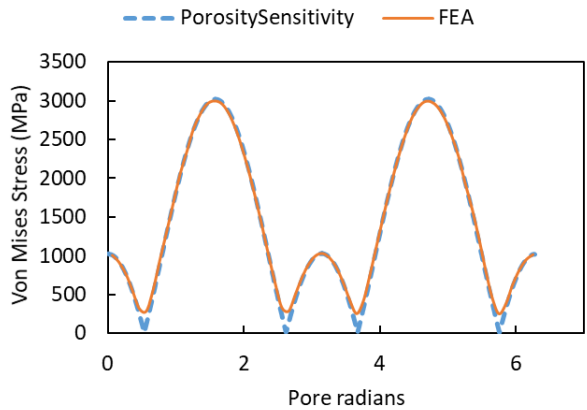

*Figure 26: Comparison of radian (θ) von Mises stress on pore surfaces between the proposed estimator and FEA.*

In this section, the same 2D beam is utilized where it is clamped at the left side and a horizontal uniform extension load (L=1000 N/mm²) is applied on the right side. A spherical hole of size 5 mm is placed at the center of the beam. von Mises stress on the pore surface approximated by exterior approximation is compared with FE analysis solutions. As illustrated in Figure 26, the approximated loop stress is highly accurate, especially on the highest stressed location with an error smaller than 1%.

In this section, we used benchmark examples to illustrate the influence of various pore parameters on the accuracy of the proposed porosity sensitivity-based estimator. It is shown to be robust and accurate, and therefore extended to the following case studies.

### 6.2 Case study: 2D control arm

A control arm was first designed in [58] through topology optimization for minimizing weight with a strength requirement. Its optimized geometry is shown in Figure 27(a). If the control arm could be manufactured by metal casting, the casting process could be simulated in MAGMASOFT [59] as in Figure 27(b) where two feeders are placed on each side.

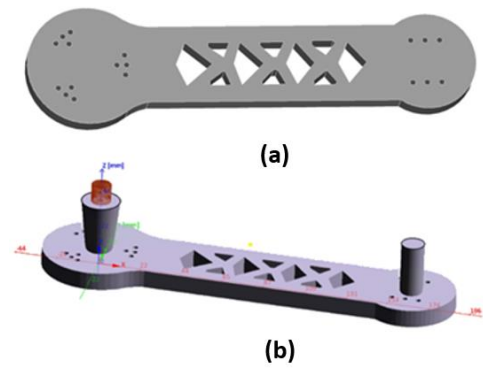

*Figure 27: **(a)** 3D CAD model of the control arm regenerated from [58], and **(b)** its casting simulation setup in MAGMASOFT.*

A top view of simulated pore distribution from MAGMASOFT is illustrated in Figure 28. It is observed the pores are mostly located close to: (1) bottom of feeders where metals are last to solidify, (2) top and bottom surfaces where multiple solidification fronts could occur, and (3) geometry intersections where two liquid metal flows meet.

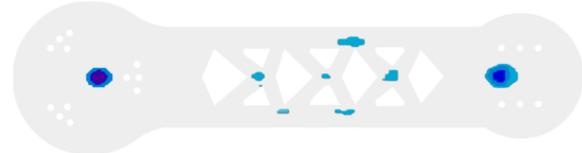

*Figure 28: Top view of simulated pore distribution via MAGMASOFT.*

To compare the proposed porosity sensitivity-based estimator with ground truth FE results, a porous geometry needs to be created, on which a direct FE analysis can be applied. Since the thickness dimension of the control arm is much less than the other two and the directions of applied loads are assumed not in thickness direction, a 2D plane stress representation could be sufficient in the FE analysis. Also, because the real shapes of pores cannot be predicted via MAGMASOFT simulation, 36 pores of different sizes and shapes, examined by light microscope in [60], are regenerated and placed in locations where pore distribution indicates, as in Figure 29. The porous geometry is then meshed using 95,888 linear triangle elements.

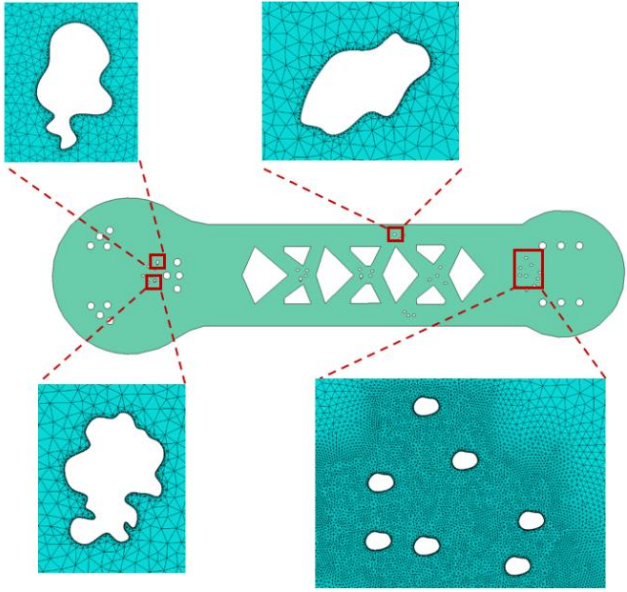

*Figure 29: 2D plane stress representation of the control arm with meshed pore shapes regenerated from [60].*

For this numerical study, the control arm model is clamped at the fixture holes on the left side, while horizontal loads (F1=6000 N) and vertical loads (F2=3000 N) are applied at the loading holes on the right side, as shown in Figure 30(a). The proposed porosity sensitivity is computed on the reference dense geometry in Figure 30(b). The quantities of interest studied include global elastic compliance, local averaged horizontal displacement in a (yellowed marked) region of interest ($\Omega_s$), and a vertical nodal displacement at a (red marked) nodal point ($P_t$) in Figure 30(b).

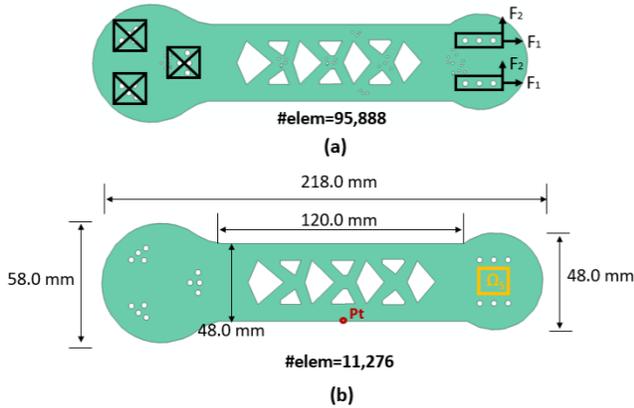

*Figure 30: (a) Boundary conditions, and (b) dimensions and regions of interests for the 2D control arm model.*

The estimation results are compared against FE analysis in Table 2 where effectivity indices are measured in terms of both performance function in Equation (6.1) and the error in Equation (6.2). It is observed, compared with FE benchmark, the proposed estimator is fairly accurate with the effectivity indices ($I_\Psi$) within [0.997, 1.004] and effectivity indices ($I_D$) in the range of [0.732, 1.065]. Comparably, the prediction error on the regional averaged displacement is higher. One plausible reason could be that in the porous geometry (Figure 30(a)), the region of interest ($\Omega_s$) is in close vicinity to the region of pores where strain concentration may occur. Strain concentration induced high displacement gradients could be problematic for the proposed first order estimation model.

*Table 2: Result comparison between direct FEA and the proposed estimator for the control arm model.*

|  | FEA | Porosity Sensitivity | Effectivity index ($I_\Psi$) | Effectivity index ($I_D$) |
|---|---|---|---|---|
| Compliance (N*mm) | 16.207 | 16.199 | 1.0004 | 1.065 |
| TipDisp (mm) | 1.205e-4 | 1.200e-4 | 1.004 | 1.052 |
| RegDisp (mm) | 8.471e-6 | 8.496e-6 | 0.997 | 0.732 |

In order to evaluate the accuracy of the approximated stress fields near pores, a histogram of stress effectivity indices at each pore is plotted in Figure 31, where the effectivity index is defined as the ratio between the maximum von Mises stress from FE analysis and the value from exterior approximation.

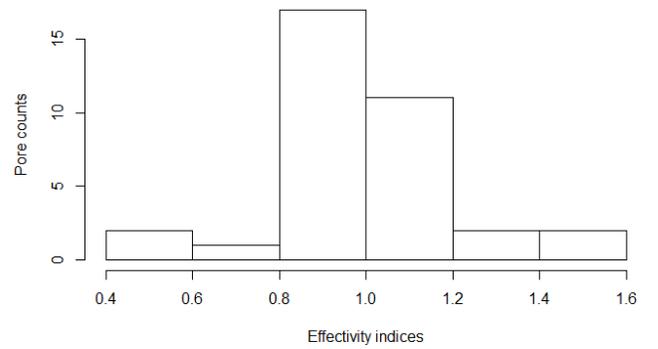

*Figure 31: Histogram of effectivity indices for von Mises stress approximations on pore surfaces in control arm.*

It can be observed from Figure 31 that the effectivity indices for the 29 out of 36 pores are in the range [0.8, 1.2], indicating the exterior formulation on stress approximations is sufficiently accurate for the majority of pores.

In addition, while the highest von Mises stress from FE analysis occurs on the pore surface in the lower middle part and close to bottom surface (Figure 32) with the value of 8.059e4 MPa, exterior approximation correctly predicts the location of the highest von Mises with a value of 7.781e4 MPa, i.e., an error smaller than 3.5%.

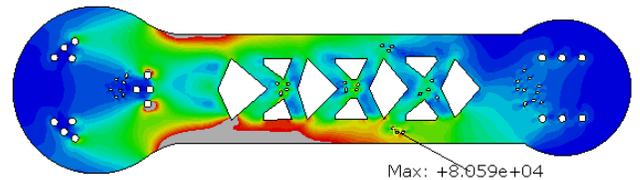

*Figure 32: von Mises stress plot of the flawed control arm model with the highest stress occurring on one of its pore boundaries.*

### 6.3 Case study: 3D tensile bar

In order to study the performance of the porosity sensitivity-based estimator on 3D modeling, we first apply it to a tensile bar which was manufactured, and CT scanned by VALUCT (250 µm resolution) [61]. Porosity detection was based on thresholding and segmented phase contrast of image processing which were performed by Image Processing Toolbox in MATLAB [62]. The detected pore geometry was then saved in STL format whose mesh was then repaired in SALOME [63] to remove mesh defects, e.g., disconnected geometry, mis-oriented face, intersected elements, etc. After mesh clean-up, a surface mesh representation of the model is seen in Figure 33.

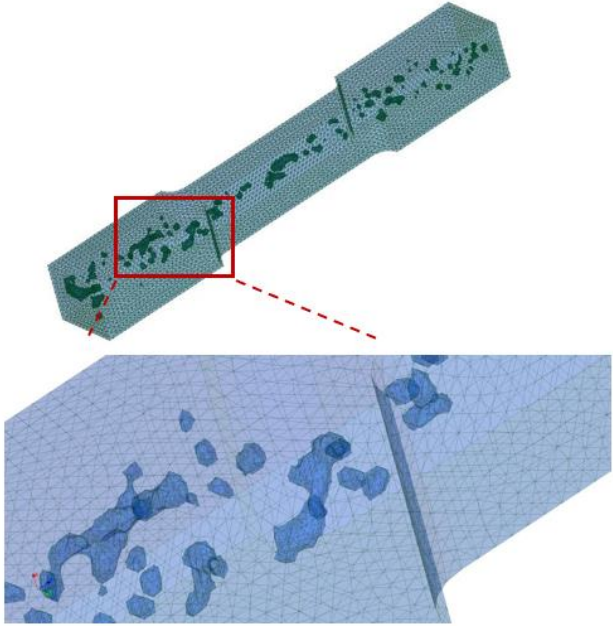

*Figure 33:* Surface mesh representation of the 3D tensile bar with internal pores.

It could be observed in Figure 33 that there are 92 internal pores whose equivalent sizes range between 0.25 mm and 1.65 mm. The pores with spherical shapes could come from entrapped gas, and the large irregular ones are more likely due to solidification. It is also seen that most of pores are distributed on the middle plane along the thickness direction. The locations of pores generally indicate where the liquid metal is last to be solidified while their surrounding solidified metals tends to shrink away. The surface mesh representation in Figure 33 is used to generate a volume mesh of 414,038 linear tetrahedral elements. It is noted the linear elements were automatically generated by converting triangle mesh (STL) to tetrahedral mesh in ABAQUS. Higher order elements (if possible) are suggested for better stress estimations. This volume mesh representation of the porous tensile bar is now ready for a FE analysis whose result as a ground truth will be compared with the proposed model.

The porosity sensitivity-based estimator is computed on the reference dense model in Figure 34 which is generated in SOLIDWORKS [64]. To simulate a tension test, all degrees of freedom (DOF) on the top and bottom surfaces of gripping sections are fixed, and a uniform surface traction (F=100,000 N/mm2) is utilized as the tensile load as seen in Figure 34. The predicted performance functions include the structural compliance, an averaged displacement in a (yellow marked) local region $\Omega_s$, and a nodal displacement at the (red marked) point ($P_t$). The model is then meshed by 17,287 linear tetrahedral elements, accounting for only 4.2% of the mesh size of the porous model solved by the direct FE approach.

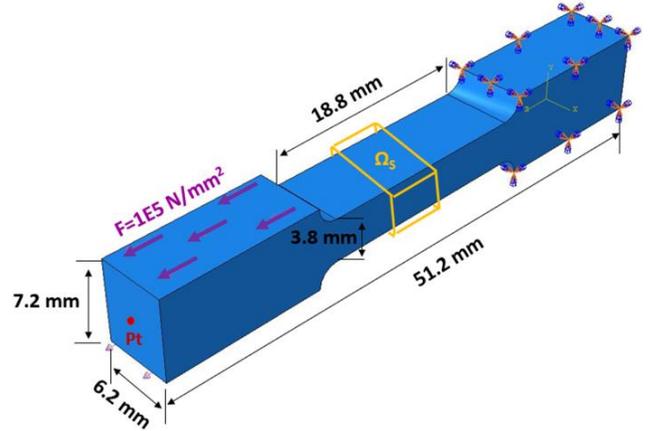

*Figure 34:* Dimensions, boundary conditions, and regions of interests on the 3D tensile bar.

The comparison results between a direct FE approach and the proposed estimator are summarized in Table 3 where the effectivity indices are 0.998 for $I_\Psi$ in Equation (6.1) or in range [0.899, 0.914] for $I_D$ in Equation (6.2), both of which indicate fairly good predictions.

*Table 3:* Result comparison between direct FEA and the proposed estimator for the tensile bar model.

|  | FEA | Porosity Sensitivity | Effectivity index ($I_\Psi$) | Effectivity index ($I_D$) |
|---|---|---|---|---|
| Compliance (N*mm) | 2.779e3 | 2.782e3 | 0.998 | 0.914 |
| TipDisp (mm) | 3.149e-4 | 3.155e-4 | 0.998 | 0.885 |
| RegDisp (mm) | 1.431e-6 | 1.433e-6 | 0.998 | 0.899 |

To compare stress fields, for the direct FE analysis, its highest von Mises stress is detected on the pore surface located in the middle gauge section (Figure 35) with a value of 1.782e6 MPa. The proposed exterior stress approximation indicates the same maximum stress location with an estimation value of 2.057e6 MPa, i.e., a 13.4% overshoot error.

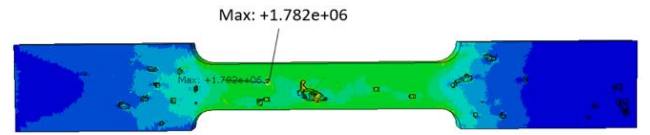

*Figure 35:* Cross-section view of von Mises stress on the porous tensile bar model with the highest stress occurring on one of its pore boundaries.

A histogram is generated to summarize the accuracy of stress approximations in all pores in Figure 36. It could be observed the majority (i.e., 82 out of 92 pores) of stress approximations is fairly accurate, with the effectivity indices ranging between 0.7 and 1.2. This suggests that the proposed method is fairly accurate for a simple 3D tensile bar model with less than 100 pores.

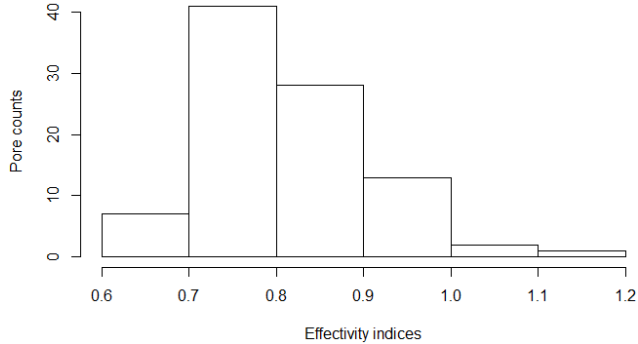

*Figure 36: Histogram of effectivity indices for von Mises stress approximations on pore surfaces in tensile bar.*

The merits of the proposed method can be emphasized from two perspectives. First, in FEM, creation of a high-quality mesh is prerequisite, and in an ideal scenario, the meshing process can be fully automated without human intervention [65]. However, automation of meshing is challenging for complex geometries; manufacturing pores often manifest complex morphologies with disconnected topologies, concave radii and sharp corners. Without human intervention, meshing will often result in inverted or tangled elements [24]. The proposed method, on the other hand, only requires meshing a simple geometry with no pores, and this can be easily automated. Second, during the solving process, we use the boundary element method [51] to include the effects of the pores. This requires a smaller memory footprint, and therefore faster execution. Table 4 compares the time taken in seconds, for various steps in direct FEM, and the proposed method.

*Table 4: Time comparison (in seconds) on different steps between direct FEM and the proposed estimator.*

|                          | Direct FEA | Proposed |
| ------------------------ | ---------- | -------- |
| Pore reconstruction (sec)| 19         | 19       |
| Meshing (sec)            | 1210       | 2        |
| Solving (sec)            | 279        | 122      |
| Total time (sec)         | 1508       | 143      |

As seen from Table 4, the total time of the proposed method accounts for less than 10% of the direct FEM approach. In the proposed method, about two thirds of the total time is spent on solving the exterior formulations of Equation (5.28) and (5.31) by boundary element method in a sequential manner; this can further be accelerated through parallelization.

**6.4 Case study: 3D W-profile plate**

The porosity properties in the W-profile plate have already been discussed in Section 2. In this section, the plate model is studied to illustrate the accuracy and robustness of the proposed method.

Different from the tensile bar model which only contains 92 pores, the number of pores in the W-profile plate is much higher at 1320. The usual image processing procedure is first used to detect and compute geometries of every pore, and then store them as surface mesh in a single STL file, as described in the 3D tensile bar section. However, due to the large number of included pores in the W-plate model, the generated STL file is so large that it would be computational prohibitive to repair mesh defects or convert to a volume mesh in meshing software, e.g., SALOME. An alternative approach is to use spherical pore models with consistent volumes to represent the real pore geometries. The porous model is therefore generated as in Figure 37, on which a direct FE analysis is applied, and its results will be compared with the porosity sensitivity-based method in the following.

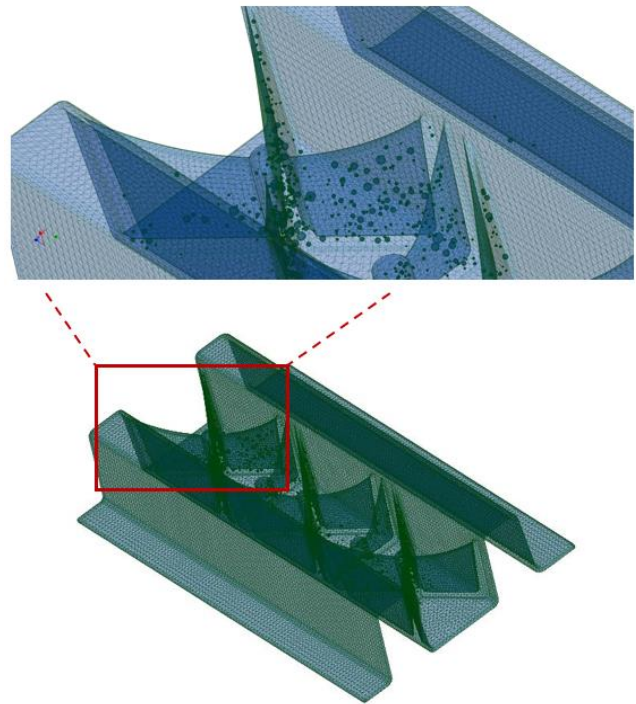

*Figure 37: Surface mesh representation of the W-profile plate with internal spherical shaped pores.*

The major advantage of using computer generated pore model to its real geometry is that its model parameters can be easily varied. In order to preserve the relation between pore volumes and sphericities in Figure 4, a 3D model [13] is created for representation of each pore. As shown in Figure 38, this model is generated by intersecting 3 identical ellipsoids at their geometry centers. While its volume is consistent with the original pore, its major and minor axis lengths are calibrated so that its sphericity is also preserved. This synthetic pore model is then used in the proposed porosity sensitivity approach.

The reason for this pore model not used in the direct FE model is for the following reason. As shown in Figure 4, a large pore tends to be long and slim shaped, and so is the shape of its pore model. While as shown in Figure 37 the W-plate is a thin wall structure, the original long slim shaped pore orients along the direction of thin walls. However, since there is no available information about pore orientations in tomography scanning data, placing a pore model (surface mesh in STL format) of low sphericity inside a thin wall geometry (surface mesh in STL format) could lead to scenarios where the triangles defined in STLs intersect with each other, resulting to mesh defects.

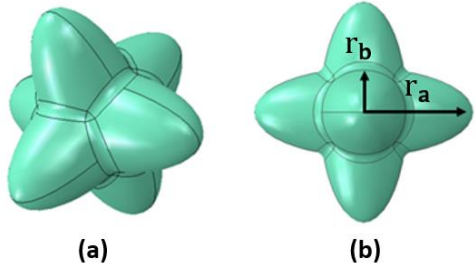

*Figure 38:* **(a)** 3D pore model [66], and **(b)** its major and minor semi axis.

If we use the pore model to represent the original pores in Figure 4, the correlation between its aspect ratio (a/b) and major axis length could be described by a scatter plot as shown in Figure 39. Consistency could be found between Figure 4 and Figure 39. For example, a hydrogen induced gas pore with a high sphericity value and small in volume ( Figure 4) could be described by a low aspect ratio (a/b) and a small major semi-axis in Figure 39.

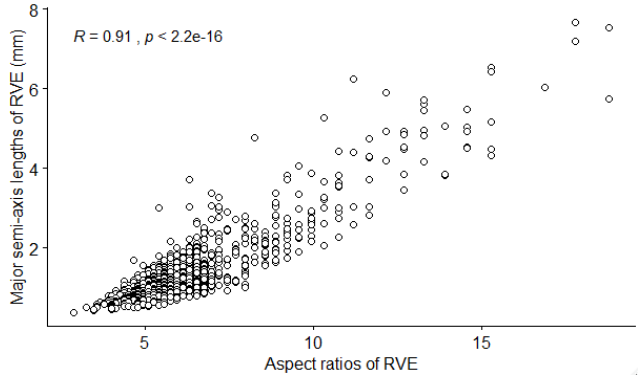

*Figure 39:* Correlation between major semi-axis length and aspect ratio on synthetic pore models from tomography data.

If all the pores in the W-plate could be represented by the pore model in Figure 38, the proposed estimator could then be applied in a similar procedure as in the study of 3D tensile bar. However, in a real situation, the number of pores in a manufactured part could be so large that even storing every pore model could be memory demanding. Therefore, in order to reduce computational cost in such scenario, instead of saving every pore model, a collection of samples is generated to statistically represent its distribution. The first step is to study the pore model property distribution from population.

In order to characterize the distributions of pore model's aspect ratio and major semi-axis length, 5 different statistical distributions are compared, as in Table 5 and

Table 6. It can be observed from the two tables that the log normal distribution gives the best fit for both aspect ratio and major semi-axis length. This is because by comparing with other distributions, log normal give the highest value of log likelihood and the lowest values of both Akaike information criterion (AIC) and Bayesian information criterion (BIC).

*Table 5:* Comparison of different statistical distributions of data fitting of pore model's major axis length.

|  | Log likelihood | AIC | BIC |
|---|---|---|---|
| Normal | -2769.513 | 5543.026 | 5553.394 |
| Gamma | -2517.897 | 5039.793 | 5050.161 |
| Weibull | -2792.490 | 5588.980 | 5599.347 |
| Lognormal | -2423.443 | 4850.886 | 4861.254 |
| Exponential | -3667.927 | 7337.854 | 7343.038 |

*Table 6:* Comparison of different statistical distributions on data fitting of pore model's aspect ratios

|  | Log likelihood | AIC | BIC |
|---|---|---|---|
| Normal | -1768.133 | 3540.267 | 3550.634 |
| Gamma | -1254.390 | 2512.780 | 2523.148 |
| Weibull | -1391.951 | 2787.902 | 2798.270 |
| Lognormal | -1098.491 | 2200.983 | 2211.350 |
| Exponential | -1617.175 | 3236.351 | 3241.534 |

Also, since a strong linear correlation (R=0.91) could be observed between the pore model aspect ratio and major semi-axis length in Figure 39, the relation between the two parameters is simplified to a linear regression. Therefore, based on the linear dependency, we generate 50 samples on the value of aspect ratios, their linearly dependent major semi-axis lengths would also satisfy a log normal distribution. The sampling of the two pore model parameters is illustrated as in Figure 40.

A norm distance between a sampling pore model and an arbitrary pore model in population (in Figure 39) could be defined as

$$d = \|P_i - S_j\|_2, \ (P_i, S_j) \in (Aspect\ Ratio, Major\ SemiAxis) \quad (6.3)$$

where $P_i$, ($i$=1, 2, ..., number of pores in casting) refers to a (pore) point in Figure 39 and $S_j$ ($j$=1,2, ..., 50) is one pore model out of 50 samples. For every population point $P_i$, if a sampling point $S_j$ could be found such that the distance $d$ is minimized, the pore at $P_i$ could then be represented by the pore model at $S_j$. The pore models are therefore ready for the proposed porosity sensitivity analysis.

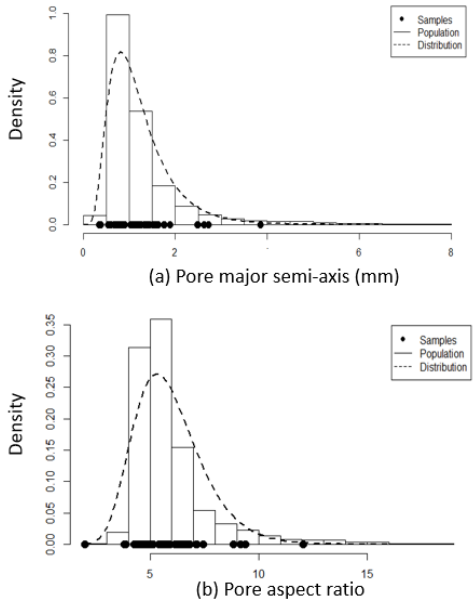

Figure 40: (a) Density distribution of pore model's major axis lengths, and (b) Density distribution of aspect ratios and 50 sampled points.

For this numerical experiment, the W-plate is clamped at two bottom surfaces and a uniform pressure (P=1e6 N/mm$^2$) is applied on the top surface as show in Figure 41. The quantities of interests include elastic compliance, regional averaged displacement in region ($\Omega_s$), and nodal displacement at point ($P_t$).

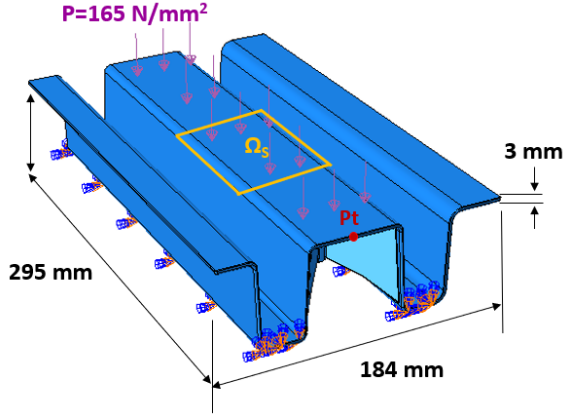

Figure 41: Dimensions, boundary conditions, and regions of interests on w-profile plate model.

The results of quantities of interests computed from the proposed estimator and a direct FE analysis are compared in Table 7. It is noted the effectivity indices of the proposed estimator in this W-plate model is less accurate than previous models. Plausible reasons could be: (1) the pore model used in direct FE analysis (spherically shaped) is different from that in Figure 38 which is used in the proposed estimator; (2) instead of using distinct pore model to represent different pore geometries, we select 1 out of 50 sampled pore models to represent each pore for the sake of computational cost; (3) the W-plate is a thin-wall structure. Included pores could be either too close to the structure surface (Figure 3(c)) or to other pores (Figure 3(d)). Pore-to-pore or pore-to-surface interactions could therefore deteriorate estimation accuracy since they are neglected in the proposed first order sensitivity field; (4) when the number of pores is large, the error of the porosity sensitivity estimator on each pore could be accumulated. But it is also worth notice that the effectivity index (i.e., 3.435) of the regional averaged displacement is still considered as acceptable and useful in the field of goal-oriented estimations [31].

Table 7: Result comparison between direct FEA and the proposed estimator for W-profile plate model.

|  | FEA | Porosity Sensitivity | Effectivity index ($I_\Psi$) | Effectivity index ($I_D$) |
|---|---|---|---|---|
| Compliance (N*mm) | 1.319e8 | 1.308e8 | 1.008 | 2.230 |
| TipDisp (mm) | 6.332e-2 | 6.334e-2 | 0.999 | 0.889 |
| RegDisp (mm) | 1.383e-2 | 1.414e-2 | 0.978 | 3.435 |

The von Mises stress on pore surfaces is compared between the exterior approximation and direct FE analysis results, as in Figure 42. The value of the highest von Mises stress is 3.774e7 MPa from FE analysis, and 4.647e7 MPa from exterior approximation on the same pore, i.e., an 18% error. Observe that in Figure 42 the average of effectivity index is around 0.7, which means for most pores the exterior approximation overestimates its stress by about 30%. The overestimation could come from the fact that different synthetic pore models are used in direct FE analysis and the proposed estimator. Compared with spherical pores utilized in direct FE analysis, the stress concentration factor on the geometry of pore model in Figure 38 is much higher, thus leading to the overshot estimation.

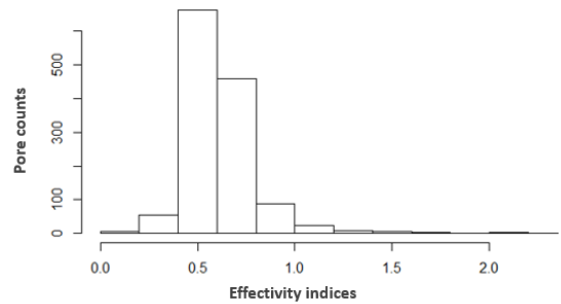

Figure 42: Histogram of effectivity indices for von Mises stress approximations on the surfaces of synthetic pore models in Figure 38.

In order to further investigate the reasons of relatively large estimation errors on stresses near pore surfaces, the porosity sensitivity estimator was computed with 1320 distinct spherical pore models (as opposed to mapping these pores to 50 template pore models). The resulting stress effectivity indices is illustrated in Figure 43; the average effectivity index is 1.092 (i.e., closer to 1.0). Comparing Figure 43 and Figure 42, the improvement on stress approximation is significant due to the consistent pore models. However, in scenarios where pore morphologies are more complex than spheres, sophisticated pore models are suggested [66].

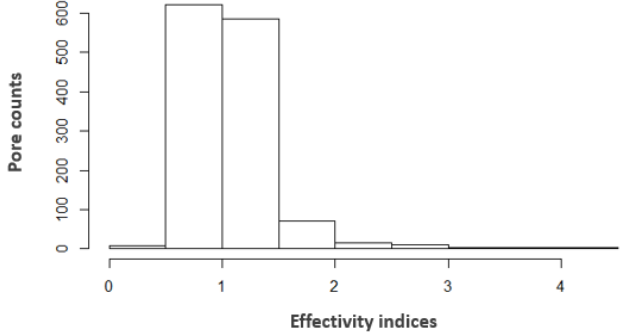

*Figure 43: Histogram of effectivity indices for von Mises stress approximations on the surfaces of spherical pore models.*

In this section, the accuracy and robustness of the proposed porosity sensitivity were tested via different numerical experiments. It was tested on a 2D benchmark example where the influence of various pore properties was explored. The proposed estimator is accurate and robust to different conditions of pore shape, size, clustering, or distance to part surface. The method is then applied to 3 different cases. For the 2D control arm and 3D tensile bar, since the number of pores in both models are small, pore geometries could be reconstructed from scanning data and saved in a lightweight STL file. On the other hand, the W-profile plate, contains too many pores that it would be computational prohibitive to regenerate and store the geometries for each pore. After studying the distributions of pore parameters from scanning data, a sampling technique is utilized to generate a pool of pore models from which the pore geometries could be statistically represented. The case studies confirm that the proposed methodology provides a reliable and computational efficient alternative to a direct FE analysis.

## 7. CONCLUSIONS

A porosity sensitivity model has been developed to quantify linear elastic quantities for cast components containing internal pores. Without explicitly solving FE problems on porous structures, the porosity sensitivity approach provides reliable estimations on quantities of interests by combining topological and shape sensitivity fields. Specifically, topological sensitivity is utilized to capture the first order change when an infinitesimally small hole is inserted into structures, and an integrated shape sensitivity field approximates the change when the small hole is continuously perturbed to resemble the shape of an arbitrary pore in the manufactured part. Even though the proposed methodology is developed, applied and verified by HPDC related porosity problems, its extension to other manufacturing (e.g., additive manufacturing [67], [68]) induced porosity problems is possible but needs to be established.

Future work on this topic includes the following. First, interactions between different pores become significant when their distance is close. Future work will include the 2nd order interaction terms to account for this effect [32]. Second, fatigue analysis on porous structures is arguably more important [19], [60]. This work could be extended to nonlinearity. Third, material properties in this paper are assumed to be isotropic. This assumption could be relaxed in future work to consider a more realistic scenario for manufactured alloys with microstructures and multiple materials with non-isotropic properties [69]. Fourth, compared with pores in casting aluminum alloys, the pores in additive manufactured alloys (e.g., titanium) are much smaller [67], [68]. Even though the proposed method was tested for various pore morphologies and spatial distributions, to account for additive manufactured induced pores, its prediction accuracy on small pores still needs to improve. Fifth, other interesting topics may include extension to multi-physics problems [70].

## ACKNOWLEDGEMENTS

The authors thank the ACRC consortium members for funding the project and also thank the industrial members of the focus group whose help was invaluable. Specifically, we appreciate Randy Beals from Magna International provided the W-profile plate for testing, and Chen Dai from VJ Technologies provided support on X-ray computed tomography (CT) scanning and data generation. NSF Grant CMMI-1824980 is also acknowledged. The authors greatly appreciate the reviewers that helped us clarify and improve the manuscript.

## Appendix A. Weak formulation

For a linear elastic problem, its strong formulation could be written as [25]

$$\begin{cases} -\nabla \cdot \sigma(z) = f^b & x \in \Omega \\ z = \hat{z} & x \in \Gamma^h \\ \sigma(z) \cdot n = f^s & x \in \Gamma^s \end{cases} \quad (A.1)$$

where $z$ is the displacement vector to be computed, $\sigma(z)$ is the stress tensor, $f^b$ is the body force for a material point $x$ defined in the domain $\Omega$, $f^s$ is the surface traction prescribed on Neumann boundary $\Gamma^s$, $n$ is the unit outward normal direction vector on $\Gamma^s$, $\Gamma^h$ is the Dirichlet boundary.

Its solution space could be defined as

$$D_A = \left\{ z \in C^2(\Omega) \mid z = \hat{z} \text{ on } x \in \Gamma^h, \sigma \cdot n = f^s \text{ on } x \in \Gamma^s \right\} \quad (A.2)$$

For a linear elastic material, its constitutive equation is given as

$$\begin{cases} \sigma = C\varepsilon \\ \varepsilon = \frac{1}{2}(\nabla z + \nabla z^T) \end{cases} \quad (A.3)$$

with its strain energy as

$$U(z) = \frac{1}{2} \iiint_\Omega \sigma(z) : \varepsilon(z) d\Omega \quad (A.4)$$

and its external work is given as

$$W(z) = \iiint_\Omega z^T f^b d\Omega + \iint_{\Gamma^s} z^T f^s d\Gamma \quad (A.5)$$

The total potential energy is defined as

$$\Pi(z) = U(z) - W(z) \quad (A.6)$$

A stable displacement solution could be achieved by minimizing the total potential energy through variational approach [25], [29]

$$\delta \Pi(z,\bar{z}) = 0 \qquad \bar{z} \in Z \tag{A.7}$$

where

$$\delta U(z,\bar{z}) = \iiint_{\Omega} \sigma(z):\varepsilon(\bar{z})d\Omega \equiv a(z,\bar{z}) \tag{A.8}$$

$$\delta W(z,\bar{z}) = \iiint_{\Omega} \bar{z}^T f^b d\Omega + \iint_{\Gamma^s} \bar{z}^T f^s d\Gamma \equiv l(\bar{z}) \tag{A.9}$$

where the virtual arbitrary field $\bar{z}$ is in a Hilbert space satisfying kinematically admissible displacements

$$Z = \{z \in H^1(\Omega) \mid z = 0 \text{ on } \Gamma^h\} \tag{A.10}$$

Then the weak formulation of Equation (A.1) can be expressed as finding a solution $z \in Z$ to

$$a(z,\bar{z}) = l(\bar{z}) \tag{A.11}$$

**Appendix B. Material derivative**

As described in Section 3, if $\tau$ could represent an infinitesimally small shape perturbation in the vicinity of a specific value $\eta$, i.e., $\tau = d\eta$, then a pointwise material derivative of displacement fields during the perturbation could be defined as [29]

$$\dot{z} = \lim_{\tau \to 0} \left[ \frac{z_\tau(x + \tau V(x)) - z(x)}{\tau} \right] \tag{B.1}$$
$$= z'(x) + \nabla z V(x) \qquad \dot{z} \in Z$$

where the spatial derivative is defined as

$$z' = \lim_{\tau \to 0} \left[ \frac{z_\tau(x) - z(x)}{\tau} \right] \tag{B.2}$$

If a domain functional could be defined as an integral on the perturbed domain,

$$\Psi = \iiint_{\Omega_\tau} f_\tau(x_\tau) d\Omega_\tau \tag{B.3}$$

where $f_\tau(x_\tau)$ is a scalar function defined on $\Omega_\tau$,

then its material derivative could be expressed as

$$\Psi' \equiv \frac{d\Psi}{d\eta} = \iiint_{\Omega} f'(x) d\Omega + \iint_{\Gamma} f(x) V_n d\Gamma \tag{B.4}$$

wherein $V_n$ is the normal component of design speeds on perturbed boundaries.

Similarly, for a functional defined over a perturbed boundary surface $\Gamma_\tau$,

$$\Psi = \iint_{\Gamma_\tau} f_\tau(x_\tau) d\Gamma_\tau \tag{B.5}$$

its material derivative could be expressed as

$$\Psi' \equiv \frac{d\Psi}{d\eta} = \iint_{\Gamma} \left[ f'(x) + (\nabla f^T n + \mathrm{K} f(x)) V_n \right] d\Gamma \tag{B.6}$$

wherein K is the mean of the surface curvature.

**Appendix C. Adjoint formulations**

In linear elastic problems, a generic quantity of interests could be defined as

$$\Psi = \iiint_{\Omega_s} g(x) d\Omega \tag{C.1}$$

where $g$ is an arbitrary function defined over a region of interests $\Omega_s$.

For the sake of simplicity, we have the following assumption in this paper. First, only quantities of interests which directly depend on displacement fields are considered, for example elastic compliance, local averaged displacement, and nodal displacement. The issue about extension to include displacement gradient dependent quantity, e.g., averaged stress components or Von Mises stress, will be discussed in the end of this section; second, the regions of interests, over which quantities of interests are computed, are assumed fixed. Therefore, adjoint formulations of the aforementioned quantities could be derived as follows.

Let us assume both body force and surface tractions are independent of shape perturbations,

$$f_b' = 0 \tag{C.2}$$

$$f_s' = 0 \tag{C.3}$$

invoking Equation (B.4) and Equation (B.6) to take material derivative of Equation (A.11) leads to

$$a'(z,\bar{z}) = l'(\bar{z}) \tag{C.4}$$

where

$$a'(z,\bar{z}) = \iiint_{\Omega} \left[ \varepsilon(\bar{z}'):\sigma(z) + \varepsilon(\bar{z}):\sigma(z') \right] d\Omega + \iint_{\Gamma} \left[ \varepsilon(\bar{z}):\sigma(z) \right] V_n d\Gamma \tag{C.5}$$

and

$$l'(\bar{z}) = \iiint_{\Omega} f_b^T \bar{z}' d\Omega + \iint_{\Gamma^{f+s}} (f_b^T \bar{z}) V_n d\Gamma + \iint_{\Gamma^s} \left\{ f_s^T \bar{z}' + \left[ \nabla(f_s^T \bar{z})^T n + \mathrm{K}(f_s^T \bar{z}) \right] \right\} V_n d\Gamma \tag{C.6}$$

Equating Equation (C.5) with (C.6), we could have

$$a(\dot{z},\bar{z}) = \iiint_{\Omega} \left[ \sigma(z):\varepsilon(\nabla \bar{z}^T V) + \sigma(\nabla z^T V):\varepsilon(\bar{z}) \right] d\Omega$$
$$- \iiint_{\Omega} f_b^T (\nabla \bar{z}^T V) d\Omega - \iint_{\Gamma} \left[ \varepsilon(\bar{z}):\sigma(z) \right] V_n d\Gamma$$
$$+ \iint_{\Gamma^{f+s}} \left[ f_b^T \bar{z} \right] V_n d\Gamma$$
$$+ \iint_{\Gamma^s} \left\{ -f_s^T (\nabla \bar{z}^T V) + \left[ \nabla(f_s^T \bar{z})^T n + \mathrm{K}(f_s^T \bar{z}) \right] V_n \right\} d\Gamma \tag{C.7}$$

where both $\dot{z}$ and $\bar{z}$ satisfy the kinematic admissible boundary condition in Equation (A.10).

<u>Compliance</u>

With the elastic compliance written as [29]

$$\Psi_J = \iiint_{\Omega} z^T f^b d\Omega + \iint_{\Gamma^s} z^T f^s d\Gamma \tag{C.8}$$

its material derivative could be derived by invoking Equation (B.4) and Equation (B.6) as

$$\Psi_J' = \iiint_{\Omega} f_b^T (\dot{z} - \nabla z V) d\Omega$$
$$+ \iint_{\Gamma} \left[ \begin{array}{c} f_b^T z V_n + f_s^T \dot{z} - f_s^T \nabla z V \\ + (\nabla(f_s^T z))^T n + \mathrm{K}(f_s^T z) V_n \end{array} \right] d\Gamma \tag{C.9}$$

Since $\dot{z}$ is in Hilbert space, it could be replaced by a virtual arbitrary displacement $\bar{\lambda}$ on right hand side of Equation (C.9) and by equating with energy bilinear term in Equation (A.8), we could arrive at the following adjoint formulation in a weak form

$$a(\lambda, \bar{\lambda}) = \iiint_\Omega f_b^T \bar{\lambda} d\Omega + \iint_\Gamma f_s^T \bar{\lambda} d\Gamma \quad (C.10)$$

where

$$\bar{\lambda} \in Z \quad (C.11)$$

The adjoint could also be written in a strong form in the following manner

$$\begin{cases} -\nabla \cdot \sigma(\lambda) = f^b & x \in \Omega \\ \lambda = 0 & x \in \Gamma^h \\ \sigma(\lambda) \cdot n = f^s & x \in \Gamma^s \end{cases} \quad (C.12)$$

By comparing Equation (C.12) with (A.10) and (A.11), it is noted the adjoint solution is identical with its primary weak-form counterpart, i.e., self-adjoint for the case of elastic compliance.

Replacing $\bar{\lambda}$ with $\dot{z}$ in Equation (C.10) leads to

$$a(\lambda, \dot{z}) = \iiint_\Omega f_b^T \dot{z} d\Omega + \iint_\Gamma f_s^T \dot{z} d\Gamma \quad (C.13)$$

Similarly, by replacing $\bar{z}$ with $\lambda$ in Equation (C.7), we have

$$\begin{aligned} a(\dot{z}, \lambda) = & \iiint_\Omega \left[ \sigma(z) : \varepsilon(\nabla \lambda^T V) + \sigma(\nabla z^T V) : \varepsilon(\lambda) \right] d\Omega \\ & - \iiint_\Omega f_b^T (\nabla \lambda^T V) d\Omega - \iint_\Gamma \left[ \varepsilon(\lambda) : \sigma(z) \right] V_n d\Gamma \\ & + \iint_{\Gamma^{f+s}} \left[ f_b^T \lambda \right] V_n d\Gamma \\ & + \iint_{\Gamma^s} \left\{ -f_s^T (\nabla \lambda^T V) + \left[ \nabla (f_s^T \lambda)^T n + K(f_s^T \lambda) \right] V_n \right\} d\Gamma \end{aligned} \quad (C.14)$$

Since the bilinear term $a(\dot{z}, \lambda)$ is symmetric on variables, by equating Equation (C.13) with (C.14) and then plugging into Equation (C.9), the shape sensitivity of elastic compliance could be written as

$$\begin{aligned} \Psi_J' = & \iiint_\Omega \begin{bmatrix} \sigma(z) : \varepsilon(\nabla \lambda^T V) + \sigma(\nabla z^T V) : \varepsilon(\lambda) \\ -f_b^T (\nabla \lambda^T V) - f_b^T (\nabla z^T V) \end{bmatrix} d\Omega \\ & + \iint_\Gamma \begin{bmatrix} f_b^T z V_n - f_s^T (\nabla z^T V) \\ + (\nabla (f_s^T z)^T n + K(f_s^T z) - \varepsilon(\lambda) : \sigma(z)) V_n \end{bmatrix} d\Gamma \\ & + \iint_{\Gamma^{f+s}} (f_b^T \lambda) V_n d\Gamma \\ & + \iint_{\Gamma^s} \left\{ -f_s^T (\nabla \lambda^T V) + \left[ \nabla (f_s^T \lambda)^T n + K(f_s^T \lambda) \right] V_n \right\} d\Gamma \end{aligned} \quad (C.15)$$

Nodal displacement

If the node of interests is fixed at a point $\hat{x} \in \Omega$, the nodal displacement could be defined as [29]

$$\Psi_{nd} = \iiint_\Omega \delta(x - \hat{x}) z d\Omega \quad (C.16)$$

By invoking Equation (B.4), its material derivative could be expressed as

$$\Psi_{nd}' = \iiint_\Omega \delta(x - \hat{x})(\dot{z} - \nabla z^T V) d\Omega \quad (C.17)$$

Similar to the compliance derivation, its adjoint formation could be written as

$$a(\lambda, \bar{\lambda}) = \iiint_\Omega \delta(x - \hat{x}) \bar{\lambda} d\Omega \quad (C.18)$$

Or, equivalently, in the form of partial differential equation through distribution theory [50]:

$$\begin{cases} -\nabla \cdot \sigma(\lambda) = \delta(x - \hat{x}) & x \in \Omega \\ \lambda = 0 & x \in \Gamma^h \\ \sigma(\lambda) \cdot n = 0 & x \in \Gamma^s \end{cases} \quad (C.19)$$

Replacing the virtual displacement $\bar{\lambda}$ by $\dot{z}$ in Equation (C.18) gives us

$$a(\lambda, \dot{z}) = \iiint_\Omega \delta(x - \hat{x}) \dot{z} d\Omega \quad (C.20)$$

Equating Equation (C.20) with Equation (C.7) and inserting resulting into Equation (C.17) results in the shape sensitivity of nodal displacement as

$$\begin{aligned} \Psi_{nd}' = & \iiint_\Omega \begin{bmatrix} \sigma(z) : \varepsilon(\nabla \lambda^T V) + \sigma(\nabla z^T V) : \varepsilon(\lambda) \\ -\delta(x - \hat{x})^T (\nabla \lambda^T V) - \delta(x - \hat{x})^T (\nabla z^T V) \end{bmatrix} d\Omega \\ & - \iint_\Gamma \left[ \sigma(z) : \varepsilon(\lambda) \right] V_n d\Gamma + \iint_{\Gamma^{f+s}} \left[ \delta(x - \hat{x})^T \lambda \right] V_n d\Gamma \\ & + \iint_{\Gamma^s} \left\{ -f_s^T (\nabla \lambda^T V) + \left[ \nabla (f_s^T \lambda)^T n + K(f_s^T \lambda) \right] V_n \right\} d\Gamma \end{aligned} \quad (C.21)$$

Averaged displacement

We are also interested in the averaged displacements over a region of interests, and it could be written as [29]

$$\Psi_{ad} = \iiint_\Omega z m_p d\Omega \quad (C.22)$$

where $m_p$ is a characteristic function which is constant within the region of interests $\Omega_s$, but zero outside, with its integration equaling to one.

With Equation (B.4), its material derivative with respect to a shape perturbation could be

$$\Psi_{ad}' = \iiint_\Omega (\dot{z} - \nabla z^T V) m_p d\Omega \quad (C.23)$$

Its adjoint could be therefore expressed in weak form as

$$a(\lambda, \bar{\lambda}) = \iiint_\Omega \bar{\lambda} m_p d\Omega \qquad \bar{\lambda} \in Z \quad (C.24)$$

or in the strong form as

$$\begin{cases} -\nabla \cdot \sigma(\lambda) = m_p & x \in \Omega \\ \lambda = 0 & x \in \Gamma^h \\ \sigma(\lambda) \cdot n = 0 & x \in \Gamma^s \end{cases} \quad (C.25)$$

Combining with Equation (C.23), its shape sensitivity could then be written in a domain integral form as

$$\begin{aligned} \Psi_{ad}' = & \iiint_\Omega \left[ \sigma(z) : \varepsilon(\nabla \lambda^T V) + \sigma(\nabla z^T V) : \varepsilon(\lambda) \right] d\Omega \\ & - \iiint_\Omega f_b^T (\nabla \lambda^T V) d\Omega - \iiint_\Omega \nabla z^T V m_p d\Omega \\ & - \iint_\Gamma \left[ \varepsilon(\lambda) : \sigma(z) \right] V_n d\Gamma + \iint_{\Gamma^{f+s}} \left[ f_b^T \lambda \right] V_n d\Gamma \\ & + \iint_{\Gamma^s} \left\{ -f_s^T (\nabla \lambda^T V) + \left[ \nabla (f_s^T \lambda)^T n + K(f_s^T \lambda) \right] V_n \right\} d\Gamma \end{aligned} \quad (C.26)$$

Averaged stress

The averaged stress over a region of interests could be defined as [29]

$$\Psi_{st} = \iiint_\Omega g(\sigma(z)) m_p d\Omega \quad (C.27)$$

where $g(\sigma)$ is a continuous differentiable stress function which could involve with Von Mises stress, stress components or any stress dependent failure criteria.

The material derivative of the averaged stress could be found as

$$\Psi_{st}' = \iiint_\Omega g_{,\sigma} \left[ \sigma(\dot{z}) - \sigma(\nabla z^T V) \right] m_p d\Omega \quad (C.28)$$

The adjoint could be therefore defined as

$$a(\lambda, \bar{\lambda}) = \iiint_\Omega g_{,\sigma} \sigma(\bar{\lambda}) m_p d\Omega \qquad \bar{\lambda} \in Z \quad (C.29)$$

By distribution theory [50], the adjoint equation could be written in a strong form as

$$\begin{cases} -\nabla \cdot \sigma(\lambda) = -\nabla \cdot (g_{,\sigma} C m_p) & x \in \Omega \\ \lambda = 0 & x \in \Gamma^h \\ \sigma(\lambda) \cdot n = (g_{,\sigma} C m_p) \cdot n & x \in \Gamma^s \end{cases} \quad (C.30)$$

For Von Mises stress $\sigma_y$, the gradient of stress function could be written as

$$g_{,\sigma} = \frac{1}{2\sigma_y} \begin{bmatrix} 2\sigma_{11} - \sigma_{22} - \sigma_{33} & 6\sigma_{12} & 6\sigma_{13} \\ 6\sigma_{12} & 2\sigma_{22} - \sigma_{11} - \sigma_{33} & 6\sigma_{23} \\ 6\sigma_{12} & 6\sigma_{23} & 2\sigma_{33} - \sigma_{11} - \sigma_{22} \end{bmatrix}$$

(C.31)

Following a similar approach as the above sections, the shape sensitivity for the averaged Von Mises stress could be as follows

$$\Psi_{st}' = \iiint_\Omega \begin{bmatrix} \sigma(z):\varepsilon(\nabla \lambda^T V) + \varepsilon(\nabla z^T V):\sigma(\lambda) \\ -f_b^T(\nabla \lambda^T V) - g_{,\sigma} \sigma(\nabla z^T V) m_p \end{bmatrix} d\Omega$$

$$- \iint_\Gamma \left[ \sigma(z):\varepsilon(\lambda) \right] V_n d\Gamma + \iint_{\Gamma^{f+s}} f_b^T \lambda V_n d\Gamma \quad (C.32)$$

$$+ \iint_{\Gamma^s} \left\{ -f_s^T(\nabla \lambda^T V) + \left[ \nabla(f_s^T \lambda) \right]^T n + \mathrm{K}(f_s^T \lambda) \right\} V_n d\Gamma$$

However, our experiments in this work show that the derivative of stress field is highly sensitive to shape perturbations. This could be due to the nonconvexity of stress fields. Therefore, the proposed method has been not applicable for stress fields so far. But its derivative is present here for the sake of completeness.

**Appendix D. Shape Sensitivity in Boundary Form**

Note that the sensitivities derived in Appendix C are all expressed in domain method which is well suited for general purpose design sensitivity analysis, especially in FE method [29]. In this section, the domain method is further transformed to boundary integrations where the shape sensitivity is derived in terms of boundary integrals.

The variational identity for the primary problem defined in Equation (A.1) could be obtained by multiplying a virtual displacement field $\bar{z} \in Z$ on both sides and integrating by parts:

$$\iiint_\Omega \sigma(z):\varepsilon(\bar{z}) d\Omega - \iiint_\Omega f_b^T \bar{z} d\Omega = \iint_\Gamma \bar{z}(\sigma(z)n) d\Gamma \quad (D.1)$$

Similarly, for adjoints defined in Equation (C.12), (C.19), (C.25) and (C.30), their variational identities could be derived in the same way.

For compliance:

$$\iiint_\Omega \sigma(\lambda):\varepsilon(\bar{\lambda}) d\Omega - \iiint_\Omega f_b^T \bar{\lambda} d\Omega = \iint_\Gamma \bar{\lambda}(\sigma(\lambda)n) d\Gamma \quad (D.2)$$

For nodal displacement:

$$\iiint_\Omega \sigma(\lambda):\varepsilon(\bar{\lambda}) d\Omega - \iiint_\Omega \bar{\lambda}^T \delta(x - \hat{x}) d\Omega$$
$$= \iint_\Gamma \bar{\lambda}(\sigma(\lambda)n) d\Gamma \quad (D.3)$$

For local averaged displacement:

$$\iiint_\Omega \sigma(\lambda):\varepsilon(\bar{\lambda}) d\Omega - \iiint_\Omega \bar{\lambda}^T m_p d\Omega = \iint_\Gamma \bar{\lambda}(\sigma(\lambda)n) d\Gamma \quad (D.4)$$

For local averaged stress:

$$\iiint_\Omega \sigma(\lambda):\varepsilon(\bar{\lambda}) d\Omega - \iiint_\Omega g_{,\sigma} \sigma(\bar{\lambda}) m_p d\Omega$$
$$= \iint_\Gamma \sigma(\lambda) n \bar{\lambda} d\Gamma - \iint_\Gamma g_{,\sigma} C m_p n \bar{\lambda} d\Gamma \quad (D.5)$$

Substitute the variational identities into the domain formulations in Equation (C.15), (C.21), (C.26) and (C.32). We further simplify the problem by assuming there is no primary body force, i.e., $f_b = 0$. After dropping all the boundary terms without design speeds, we could obtain a uniform equation for the shape sensitivity in boundary integral form as

$$\Psi' = -\iint_{\Gamma^p} V_n \left[ \sigma(z):\varepsilon(\lambda) \right] d\Gamma \quad (D.6)$$

where $\Gamma^p$ refers to the pore boundaries where design speeds are defined.

**REFERENCES**

[1] J. Collot, "Review of New Process Technologies in the Aluminum Die-Casting Industry," *Materials and Manufacturing Processes*, vol. 16, no. 5, pp. 595–617, Sep. 2001, doi: 10.1081/AMP-100108624.

[2] P. W. Cleary, J. Ha, M. Prakash, and T. Nguyen, "3D SPH flow predictions and validation for high pressure die casting of automotive components," *Applied Mathematical Modelling*, vol. 30, no. 11, pp. 1406–1427, Nov. 2006, doi: 10.1016/j.apm.2006.03.012.

[3] F. Bonollo, N. Gramegna, and G. Timelli, "High-Pressure Die-Casting: Contradictions and Challenges," *JOM*, vol. 67, no. 5, pp. 901–908, May 2015, doi: 10.1007/s11837-015-1333-8.

[4] X. P. Niu, B. H. Hu, I. Pinwill, and H. Li, "Vacuum assisted high pressure die casting of aluminium alloys," *Journal of Materials Processing Technology*, vol. 105, no. 1, pp. 119–127, Sep. 2000, doi: 10.1016/S0924-0136(00)00545-8.

[5] P. W. Cleary, J. Ha, and V. Ahuja, "High pressure die casting simulation using smoothed particle hydrodynamics," *International Journal of Cast Metals Research*, vol. 12, no. 6, pp. 335–355, May 2000, doi: 10.1080/13640461.2000.11819372.

[6] K. V. Yang, C. H. Cáceres, A. V. Nagasekhar, and M. A. Easton, "The skin effect and the yielding behavior of cold chamber high pressure die cast Mg–Al alloys," *Materials Science and Engineering: A*, vol. 542, pp. 49–55, Apr. 2012, doi: 10.1016/j.msea.2012.02.029.

[7] H. I. Laukli, *High Pressure Die Casting of Aluminium and Magnesium Alloys: Grain Structure and Segregation Characteristics*. Fakultet for naturvitenskap og teknologi, 2004. Accessed: Dec. 05, 2020. [Online]. Available: https://ntnuopen.ntnu.no/ntnu-xmlui/handle/11250/248738

[8] Ş. Yazman, U. Köklü, L. Urtekin, S. Morkavuk, and L. Gemi, "Experimental study on the effects of cold chamber die casting parameters on high-speed drilling

machinability of casted AZ91 alloy," *Journal of Manufacturing Processes*, vol. 57, pp. 136–152, Sep. 2020, doi: 10.1016/j.jmapro.2020.05.050.
[9] H. R. Ammar, A. M. Samuel, and F. H. Samuel, "Porosity and the fatigue behavior of hypoeutectic and hypereutectic aluminum–silicon casting alloys," *International Journal of Fatigue*, vol. 30, no. 6, pp. 1024–1035, Jun. 2008, doi: 10.1016/j.ijfatigue.2007.08.012.
[10] Z. Hashin and S. Shtrikman, "A variational approach to the theory of the elastic behaviour of multiphase materials," *Journal of the Mechanics and Physics of Solids*, vol. 11, no. 2, pp. 127–140, Mar. 1963, doi: 10.1016/0022-5096(63)90060-7.
[11] R. Hill, "A self-consistent mechanics of composite materials," *Journal of the Mechanics and Physics of Solids*, vol. 13, no. 4, pp. 213–222, Aug. 1965, doi: 10.1016/0022-5096(65)90010-4.
[12] T. Mori and K. Tanaka, "Average stress in matrix and average elastic energy of materials with misfitting inclusions," *Acta Metallurgica*, vol. 21, no. 5, pp. 571–574, May 1973, doi: 10.1016/0001-6160(73)90064-3.
[13] T. Taxer, C. Schwarz, W. Smarsly, and E. Werner, "A finite element approach to study the influence of cast pores on the mechanical properties of the Ni-base alloy MAR-M247," *Materials Science and Engineering: A*, vol. 575, pp. 144–151, Jul. 2013, doi: 10.1016/j.msea.2013.02.067.
[14] F. A. L. Dullien, "Characterization of Porous Media — Pore Level," in *Mathematical Modeling for Flow and Transport Through Porous Media*, G. Dagan, U. Hornung, and P. Knabner, Eds. Dordrecht: Springer Netherlands, 1991, pp. 581–606. doi: 10.1007/978-94-017-2199-8_8.
[15] R. Kumar and B. Bhattacharjee, "Porosity, pore size distribution and in situ strength of concrete," *Cement and Concrete Research*, vol. 33, no. 1, pp. 155–164, Jan. 2003, doi: 10.1016/S0008-8846(02)00942-0.
[16] R. Gerling, E. Aust, W. Limberg, M. Pfuff, and F. P. Schimansky, "Metal injection moulding of gamma titanium aluminide alloy powder," *Materials Science and Engineering: A*, vol. 423, no. 1, pp. 262–268, May 2006, doi: 10.1016/j.msea.2006.02.002.
[17] S. Leuders *et al.*, "On the mechanical behaviour of titanium alloy TiAl6V4 manufactured by selective laser melting: Fatigue resistance and crack growth performance," *International Journal of Fatigue*, vol. 48, pp. 300–307, Mar. 2013, doi: 10.1016/j.ijfatigue.2012.11.011.
[18] G. Kasperovich, J. Haubrich, J. Gussone, and G. Requena, "Correlation between porosity and processing parameters in TiAl6V4 produced by selective laser melting," *Materials & Design*, vol. 105, pp. 160–170, Sep. 2016, doi: 10.1016/j.matdes.2016.05.070.
[19] V. Prithivirajan and M. D. Sangid, "The role of defects and critical pore size analysis in the fatigue response of additively manufactured IN718 via crystal plasticity," *Materials & Design*, vol. 150, pp. 139–153, Jul. 2018, doi: 10.1016/j.matdes.2018.04.022.
[20] P. Baicchi, G. Nicoletto, and E. Riva, "Modeling the influence of pores on fatigue crack initiation in a cast Al-Si alloy," 2006.
[21] M. Ries, C. Krempaszky, B. Hadler, and E. Werner, "The influence of porosity on the elastoplastic behavior of high performance cast alloys," *PAMM*, vol. 7, no. 1, pp. 2150005–2150006, Dec. 2007, doi: 10.1002/pamm.200700159.
[22] Z. Shan and A. M. Gokhale, "Micromechanics of complex three-dimensional microstructures," *Acta Materialia*, vol. 49, no. 11, pp. 2001–2015, Jun. 2001, doi: 10.1016/S1359-6454(01)00093-3.
[23] Y. Hangai and S. Kitahara, "Quantitative Evaluation of Porosity in Aluminum Die Castings by Fractal Analysis of Perimeter," *Materials Transactions*, vol. 49, no. 4, pp. 782–786, 2008, doi: 10.2320/matertrans.MRA2007314.
[24] J. Danczyk and K. Suresh, "Finite element analysis over tangled simplicial meshes: Theory and implementation," *Finite Elements in Analysis and Design*, vol. 70–71, pp. 57–67, Aug. 2013, doi: 10.1016/j.finel.2013.04.004.
[25] R. D. Cook, *Concepts and applications of finite element analysis*. John wiley & sons, 2007.
[26] İ. Temizer and P. Wriggers, "On the computation of the macroscopic tangent for multiscale volumetric homogenization problems," *Computer Methods in Applied Mechanics and Engineering*, vol. 198, no. 3, pp. 495–510, Dec. 2008, doi: 10.1016/j.cma.2008.08.018.
[27] L. Xia and P. Breitkopf, "Recent Advances on Topology Optimization of Multiscale Nonlinear Structures," *Arch Computat Methods Eng*, vol. 24, no. 2, pp. 227–249, Apr. 2017, doi: 10.1007/s11831-016-9170-7.
[28] L. Fine, L. Remondini, and J.-C. Leon, "Automated generation of FEA models through idealization operators," *International Journal for Numerical Methods in Engineering*, vol. 49, no. 1–2, pp. 83–108, 2000, doi: 10.1002/1097-0207(20000910/20)49:1/2<83::AID-NME924>3.0.CO;2-N.
[29] K. K. Choi and N.-H. Kim, *Structural Sensitivity Analysis and Optimization 1: Linear Systems*. Springer Science & Business Media, 2006.
[30] M. P. Bendsoe and O. Sigmund, *Topology Optimization: Theory, Methods, and Applications*. Springer Science & Business Media, 2013.
[31] M. Li and S. Gao, "Estimating defeaturing-induced engineering analysis errors for arbitrary 3D features," *Computer-Aided Design*, vol. 43, no. 12, pp. 1587–1597, Dec. 2011, doi: 10.1016/j.cad.2011.08.006.
[32] M. Li, B. Zhang, and R. R. Martin, "Second-order defeaturing error estimation for multiple boundary features," *International Journal for Numerical Methods in Engineering*, vol. 100, no. 5, pp. 321–346, 2014, doi: 10.1002/nme.4725.
[33] S. H. Gopalakrishnan and K. Suresh, "A formal theory for estimating defeaturing-induced engineering analysis errors," *Computer-Aided Design*, vol. 39, no. 1, pp. 60–68, Jan. 2007, doi: 10.1016/j.cad.2006.09.006.
[34] I. Turevsky, S. H. Gopalakrishnan, and K. Suresh, "Defeaturing: A posteriori error analysis via feature sensitivity," *International Journal for Numerical Methods in Engineering*, vol. 76, no. 9, pp. 1379–1401, 2008, doi: 10.1002/nme.2345.
[35] S. Deng and K. Suresh, "Predicting the Benefits of Topology Optimization," in *Volume 2A: 41st Design Automation Conference*, Boston, Massachusetts, USA, Aug. 2015, p. V02AT03A014. doi: 10.1115/DETC2015-46349.
[36] J. Deng and W. Chen, "Concurrent topology optimization of multiscale structures with multiple porous materials under random field loading uncertainty," *Struct*

[36] *Multidisc Optim*, vol. 56, no. 1, pp. 1–19, Jul. 2017, doi: 10.1007/s00158-017-1689-1.

[37] J. Deng, J. Yan, and G. Cheng, "Multi-objective concurrent topology optimization of thermoelastic structures composed of homogeneous porous material," *Struct Multidisc Optim*, vol. 47, no. 4, pp. 583–597, Apr. 2013, doi: 10.1007/s00158-012-0849-6.

[38] "Magna International," *Magna*. https://www.magna.com

[39] "VJ Technologies – VJ Technologies is a leading global provider of X-ray inspection solutions. We apply our radioscopic digital imaging expertise to government agencies and nondestructive testing (NDT) markets throughout the world." https://vjt.com/

[40] "VGSTUDIO MAX: High-End Software for CT Data," *volumegraphics.com*. https://www.volumegraphics.com/en/products/vgstudio-max.html

[41] A. A. Novotny, R. A. Feijóo, E. Taroco, and C. Padra, "Topological sensitivity analysis for three-dimensional linear elasticity problem," *Computer Methods in Applied Mechanics and Engineering*, vol. 196, no. 41, pp. 4354–4364, Sep. 2007, doi: 10.1016/j.cma.2007.05.006.

[42] H. A. Eschenauer, V. V. Kobelev, and A. Schumacher, "Bubble method for topology and shape optimization of structures," *Structural Optimization*, vol. 8, no. 1, pp. 42–51, Aug. 1994, doi: 10.1007/BF01742933.

[43] J. Sokolowski and A. Zochowski, "On the Topological Derivative in Shape Optimization," *SIAM J. Control Optim.*, vol. 37, no. 4, pp. 1251–1272, Jan. 1999, doi: 10.1137/S0363012997323230.

[44] A. A. Novotny, R. A. Feijóo, E. Taroco, and C. Padra, "Topological-Shape Sensitivity Method: Theory and Applications," in *IUTAM Symposium on Topological Design Optimization of Structures, Machines and Materials*, Dordrecht, 2006, pp. 469–478. doi: 10.1007/1-4020-4752-5_45.

[45] S. Deng and K. Suresh, "Multi-constrained topology optimization via the topological sensitivity," *Struct Multidisc Optim*, vol. 51, no. 5, pp. 987–1001, May 2015, doi: 10.1007/s00158-014-1188-6.

[46] S. Deng and K. Suresh, "Multi-constrained 3D topology optimization via augmented topological level-set," *Computers & Structures*, vol. 170, pp. 1–12, Jul. 2016, doi: 10.1016/j.compstruc.2016.02.009.

[47] S. Deng and K. Suresh, "Topology optimization under thermo-elastic buckling," *Struct Multidisc Optim*, vol. 55, no. 5, pp. 1759–1772, May 2017, doi: 10.1007/s00158-016-1611-2.

[48] S. Deng and K. Suresh, "Stress constrained thermo-elastic topology optimization with varying temperature fields via augmented topological sensitivity based level-set," *Struct Multidisc Optim*, vol. 56, no. 6, pp. 1413–1427, Dec. 2017, doi: 10.1007/s00158-017-1732-2.

[49] D. A. Tortorelli and W. Zixian, "A systematic approach to shape sensitivity analysis," *International Journal of Solids and Structures*, vol. 30, no. 9, pp. 1181–1212, Jan. 1993, doi: 10.1016/0020-7683(93)90012-V.

[50] W. Kecs and P. Teodorescu, *Applications of the theory of distributions in mechanics*. Taylor & Francis, 1974.

[51] A. A. Becker, *The boundary element method in engineering: a complete course*. McGraw-Hill Companies, 1992.

[52] Y. J. Liu, "A fast multipole boundary element method for 2D multi-domain elastostatic problems based on a dual BIE formulation," *Comput Mech*, vol. 42, no. 5, pp. 761–773, Oct. 2008, doi: 10.1007/s00466-008-0274-2.

[53] P. Bettess, "Infinite elements," *International Journal for Numerical Methods in Engineering*, vol. 11, no. 1, pp. 53–64, 1977, doi: 10.1002/nme.1620110107.

[54] "ABAQUS/Standard User's Manual, Version 6.9. / Smith, Michael. Providence, RI : Dassault Systèmes Simulia Corp, 2009."

[55] "RStudio Team (2015). RStudio: Integrated Development for R. RStudio, Inc., Boston, MA URL http://www.rstudio.com/."

[56] S. Prudhomme, J. T. Oden, T. Westermann, J. Bass, and M. E. Botkin, "Practical methods for a posteriori error estimation in engineering applications," *International Journal for Numerical Methods in Engineering*, vol. 56, no. 8, pp. 1193–1224, 2003, doi: 10.1002/nme.609.

[57] S. M. H. Hojjatzadeh *et al.*, "Direct observation of pore formation mechanisms during LPBF additive manufacturing process and high energy density laser welding," *International Journal of Machine Tools and Manufacture*, vol. 153, p. 103555, Jun. 2020, doi: 10.1016/j.ijmachtools.2020.103555.

[58] P. Yao, K. Zhou, Y. Lin, and Y. Tang, "Light-Weight Topological Optimization for Upper Arm of an Industrial Welding Robot," *Metals*, vol. 9, no. 9, Art. no. 9, Sep. 2019, doi: 10.3390/met9091020.

[59] "Magmasoft-5.1-Tutorial," *Software training,tutorials,download,torrent*, Dec. 15, 2011. http://www.crackcad.com/magmasoft-5-1-tutorial/ (accessed Sep. 16, 2020).

[60] P. Baicchi, G. Nicoletto, and E. Riva, *Modeling the influence of pores on fatigue crack initiation in a cast Al-Si alloy*. CP2006, 2006.

[61] "ValuCT – VJ Technologies." https://vjt.com/valuct-system/ (accessed Sep. 29, 2020).

[62] "MATLAB. (2010). version 7.10.0 (R2010a). Natick, Massachusetts: The MathWorks Inc."

[63] "A. Ribes and C. Caremoli, 'Salomé platform component model for numerical simulation,' COMPSAC 07: Proceeding of the 31st Annual International Computer Software and Applications Conference, pages 553-564, Washington, DC, USA, 2007, IEEE Computer Society."

[64] "2019 SOLIDWORKS Help - Manual Download." http://help.solidworks.com/2019/english/SolidWorks/install_guide/hid_state_manual_download.htm (accessed Dec. 03, 2020).

[65] G. R. Liu, *Mesh Free Methods: Moving Beyond the Finite Element Method*, 1st edition. CRC Press, 2002.

[66] T. Taxer, C. Schwarz, W. Smarsly, and E. Werner, "A finite element approach to study the influence of cast pores on the mechanical properties of the Ni-base alloy MAR-M247," *Materials Science and Engineering: A*, vol. 575, pp. 144–151, Jul. 2013, doi: 10.1016/j.msea.2013.02.067.

[67] J. Ning, D. E. Sievers, H. Garmestani, and S. Y. Liang, "Analytical modeling of part porosity in metal additive manufacturing," *International Journal of Mechanical Sciences*, vol. 172, p. 105428, Apr. 2020, doi: 10.1016/j.ijmecsci.2020.105428.

[68] J. A. Slotwinski, E. J. Garboczi, and K. M. Hebenstreit, "Porosity Measurements and Analysis for Metal

Additive Manufacturing Process Control," *J Res Natl Inst Stand Technol*, vol. 119, pp. 494–528, Sep. 2014, doi: 10.6028/jres.119.019.

[69] A. R. Kennedy, A. E. Karantzalis, and S. M. Wyatt, "The microstructure and mechanical properties of TiC and TiB2-reinforced cast metal matrix composites," *Journal of Materials Science*, vol. 34, no. 5, pp. 933–940, Mar. 1999, doi: 10.1023/A:1004519306186.

[70] R. D. Cook, D. S. Malkus, M. E. Plesha, and R. J. Witt, *Concepts and Applications of Finite Element Analysis, 4th Edition*, 4th Edition. New York, NY: Wiley, 2001.